\documentclass[12pt]{article}

\usepackage{axodraw}

\begin{document}

\begin{titlepage}
\begin{flushright}
UFIFT-QG-05-03 \\ CRETE-05-11 \\ gr-qc/0506056
\end{flushright}

\vspace{0.5cm}

\begin{center}
\bf{Dimensionally Regulated Graviton 1-Point Function in de Sitter}
\end{center}

\vspace{0.3cm}

\begin{center}
N. C. Tsamis$^{\dagger}$
\end{center}
\begin{center}
\it{Department of Physics, University of Crete \\
GR-710 03 Heraklion, HELLAS.}
\end{center}

\vspace{0.2cm}

\begin{center}
R. P. Woodard$^{\ddagger}$
\end{center}
\begin{center}
\it{Department of Physics, University of Florida \\
Gainesville, FL 32611, UNITED STATES.}
\end{center}

\vspace{0.3cm}

\begin{center}
ABSTRACT
\end{center}
\hspace*{.3cm}
We use dimensional regularization to compute the 1PI 1-point function of
quantum gravity at one loop order in a locally de Sitter background. As
with other computations, the result is a finite constant at this order.
It corresponds to a small positive renormalization of the cosmological 
constant.

\vspace{0.3cm}

\begin{flushleft}
\end{flushleft}

\vspace{0.1cm}

\begin{flushleft}
$^{\dagger}$ e-mail: tsamis@physics.uoc.gr \\
$^{\ddagger}$ e-mail: woodard@phys.ufl.edu
\end{flushleft}

\end{titlepage}

\section{Introduction}

Since its inception, dimensional regularization \cite{HV1,BG} as been an 
extraordinarily useful technique because it preserves continuous symmetries
that do not depend upon the special properties of a certain dimension. Although
its use is ubiquitous in flat space, the technique is not so simply applied
in curved backgrounds because one must know the propagators in an arbitrary
dimension. This is no problem for determining divergences \cite{HV2,DN1,DN2,
DTN,GS} because these are universal, but it can be a real problem for 
extracting the finite parts of the 1-Particle-Irreducible (1PI) functions
which incorporate all information about a quantum field theory.

The difficulty appears even in as simple a background as de Sitter for as 
simple a diagram as the 1PI graviton 1-point function. Up to some factors 
this is the same quantity which is often termed the expectation value of
the graviton stress-tensor. It represents the back-reaction of virtual 
gravitons upon the (de Sitter) background. Ford was the first to compute 
it at one loop order in the context of a general search for secular 
infrared corrections to the effective cosmological constant \cite{LHF}. 
Owing to the inherently infrared character of the effect he sought, Ford 
worked in $D = 3 + 1$ dimensions and evaluated only a certain part of the 
graviton stress tensor canonically in a physical gauge. He found a small, 
time independent, negative shift of the cosmological constant.

Finelli, Marozzi, Vacca and Venturi have recently computed the full graviton
stress tensor using adiabatic regularization \cite{FMVV}. Although their 
result is also independent of time, they find a small positive shift of
the cosmological constant. Because they computed slightly different things
it is not clear there is any disagreement between this result and Ford's.
In any case, both effects can be absorbed into counterterms, and {\it must}
be so absorbed if the universe is to inflate at the background Hubble
constant.

The purpose of this paper is to exploit a new result for the $D$-dimensional
graviton propagator in a locally de Sitter background \cite{RPW1} to 
compute the one loop graviton 1-point function (Fig.~1) using dimensional 
regularization. The point is not to check previous results but rather to test 
the new formalism in a setting where we know what it should give: namely, a
finite, time independent renormalization of the cosmological constant. 
Although this is, of necessity, a technical paper, there are two important
physical motivations for developing the new formalism. We shall digress
briefly to explain these before becoming immersed in technicalities.

\begin{center} 
\begin{picture}(340,100)(0,0)
\Photon(60,80)(60,50){3}{3}
\PhotonArc(60,30)(20,-360,0){3}{10}
\Vertex(60,50){3}
\Text(60,40)[b]{$x$}
\Text(60,83)[b]{$\alpha\beta$}
\Photon(170,80)(170,50){3}{3}
\CArc(170,30)(20,-360,0)
\Vertex(170,50){3}
\Text(170,40)[b]{$x$}
\Text(170,83)[b]{$\alpha\beta$}
\Photon(280,80)(280,50){3}{3}
\Vertex(280,50){3}
\Text(280,40)[b]{$x$}
\Text(281,43)[b]{\LARGE $\times$}
\Text(280,83)[b]{$\alpha\beta$}
\end{picture} 
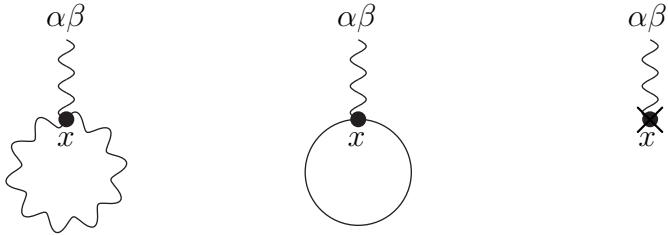
\\ {\rm Fig.~1: One loop contributions to the graviton 1-point function.} 
\end{center}

Our first motivation is to extend Ford's search past one loop order.
The genesis of back-reaction during inflation (for which de Sitter is
a paradigm) is that the expansion of spacetime continually rips long 
wavelength gravitons out of the vacuum. There is no secular back-reaction 
at one loop because the enormous growth of the total energy of these 
gravitons is canceled by the inflationary expansion of the 3-volume.
Hence the one loop energy density should amount to only a positive
constant. At the next order there must be gravitational interactions 
{\it between} the newly produced gravitons. Whereas the one loop effect 
is a constant --- which must be subsumed into a cosmological counterterm 
for the universe to begin inflation at the background Hubble constant 
--- the two loop effect should grow because each newly emerged graviton 
experiences the gravitational fields of all the gravitons produced within 
its past light-cone \cite{TW0}. A decade-old computation of the 1PI
graviton 1-point function at two loop order does indicate such an effect 
\cite{TW1} but this calculation had to be done in $D = 3 + 1$ dimensions 
using a cutoff on the co-moving 3-momentum. It is important to make the 
computation with an invariant regularization in order to ensure that 
spurious secular behavior is not being injected by what is effectively a 
time dependent ultraviolet cutoff. If the expected secular back-reaction 
occurs it could lead to a realistic model of inflation in which the
(old) problem of the cosmological constant is resolved \cite{TW2}.

Our second purpose is to facilitate a general study of graviton-mediated,
quantum effects during inflation. Massless, minimally coupled (MMC) scal\-ars
and gravitons are unique in achieving masslessness without classical
conformal invariance. This allows both particles to be produced copiously 
during inflation, which is the ultimate source of primordial cosmological
scalar \cite{MC} and tensor \cite{AAS} perturbations. It has recently been 
realized that interactions involving even a single, undifferentiated MMC
scalar can result in vastly strengthened quantum effects during inflation.
Reliable, dimensionally regulated results have been obtained in three models:
\begin{enumerate} 
\item{For a MMC scalar with a quartic self-interaction both the expectation 
value of the stress tensor \cite{OW1,OW2} and the self-mass-squared \cite{BOW} 
have been evaluated at one and two loop orders. This model shows a violation 
of the weak energy condition in which inflationary particle production drives 
the scalar up its potential and induces a curious sort of time-dependent mass.}
\item{When a complex MMC scalar is coupled to electromagnetism it has been 
possible to compute the one loop vacuum polarization \cite{PTW1,PTW2} and use 
the result to solve the quantum corrected Maxwell equations \cite{PW1}. 
Although photon creation is suppressed during inflation, this model shows 
a vast enhancement of the 0-point energy of super-horizon photons which may 
serve to seed cosmological magnetic fields \cite{DDPT,DPTD,PW2}.}
\item{When a real MMC scalar is Yukawa coupled to a massless Dirac fermion 
it has been possible to compute the one loop fermion self-energy and use it 
to solve the quantum corrected Dirac equation \cite{PW3}. The resulting 
model shows explosive creation of fermions. A recent one loop computation of 
the scalar self-mass-squared indicates that the scalar cannot develop a large 
enough mass quickly enough to prevent the super-horizon fermion modes from
becoming fully populated \cite{DW}.}
\end{enumerate}
Analogous graviton effects should be suppressed by the higher dimension of 
the respective couplings. On the other hand, they should be universal. In
particular, graviton-mediated effects do not depend upon the existence of
a minimally coupled scalar with an unnaturally light mass.

Having motivated the exercise, we close this introduction with an outline. 
In section 2 we work out the Feynman rules. The actual computation is 
described in section 3 and our conclusions are presented in section 4.

\section{Feynman Rules}

The purpose of this section is to give the Feynman rules necessary for
evaluating the diagrams of Fig.~1. We begin by expressing the invariant
action in terms of a conformally rescaled graviton field. At this point 
it is simple to read off the various 3-graviton vertex operators needed
for the first diagram. In order to get the propagators we fix the gauge
with a convenient variant of the de Donder gauge fixing term of flat
space. That determines the ghost and graviton propagators. We close with
the graviton-ghost-anti-ghost vertex operators.

In $D$ spacetime dimensions the Einstein-Hilbert Lagrangian is,
\begin{equation}
\mathcal{L} = \frac1{16 \pi G} \Bigl(R - (D\!-\!2) \Lambda\Bigr) \sqrt{-g} \; .
\end{equation}
The unique, maximally symmetric solution for positive $\Lambda$ is known as
de Sitter space. In order to regard this as a paradigm for inflation we work 
on a portion of the full de Sitter manifold known as the open conformal 
coordinate patch. The invariant element for this is,
\begin{equation}
ds^2 = a^2 \Bigl( -d\eta^2 + d\vec{x} \!\cdot\! d\vec{x}\Bigr) \qquad
{\rm where} \qquad a(\eta) = -\frac1{H\eta} \; ,
\end{equation}
and the $D$-dimensional Hubble constant is $H \equiv \sqrt{\Lambda/(D\!-\!1)}$.
Note that the conformal time $\eta$ runs from $-\infty$ to zero.

We define the graviton field $h_{\mu\nu}(x)$ as the perturbation of the
conformally rescaled metric,
\begin{equation}
g_{\mu\nu}(x) \equiv a^2 \Bigl(\eta_{\mu\nu} + \kappa h_{\mu\nu}(x)\Bigr)
\equiv a^2 \widetilde{g}_{\mu\nu} \; , 
\end{equation}
where $\kappa^2 \equiv 16 \pi G$ is the loop-counting parameter of quantum
gravity. By convention, graviton indicies are raised and lowered with the
Lorentz metric: $h^{\mu}_{~\nu} \equiv \eta^{\mu\rho} h_{\rho\nu}$,
$h^{\mu\nu} \equiv \eta^{\mu\rho} \eta^{\nu\sigma} h_{\rho\sigma}$ and $h
\equiv \eta^{\mu\nu} h_{\mu\nu}$. However, $\widetilde{g}^{\mu\nu}$ denotes
the full matrix inverse of $\widetilde{g}_{\mu\nu}$,
\begin{equation}
\widetilde{g}^{\mu\nu} = \eta^{\mu\nu} - \kappa h^{\mu\nu} + \kappa^2 
h^{\mu}_{~\rho} h^{\rho\nu} - \dots 
\end{equation}
With these conventions we can extract a surface term from the invariant
Lagrangian and write it in the form \cite{TW3},
\begin{eqnarray}
\lefteqn{\mathcal{L} \!-\! {\rm Surface} = {\scriptstyle (\frac{D}2 - 1)}
H a^{D-1} \sqrt{-\widetilde{g}} \widetilde{g}^{\rho\sigma} \widetilde{g}^{\mu
\nu} h_{\rho\sigma ,\mu} h_{\nu 0} } \nonumber \\
& & \hspace{-.7cm} + a^{D-2} \! \sqrt{\!-\widetilde{g}}
\widetilde{g}^{\alpha\beta} \widetilde{g}^{\rho\sigma} \widetilde{g}^{\mu\nu}
\Bigl\{{\scriptstyle \frac12} h_{\alpha\rho ,\mu} h_{\beta\sigma 
,\nu} \!-\! {\scriptstyle \frac12} h_{\alpha\beta ,\rho} h_{\sigma\mu ,\nu}
\!+\! {\scriptstyle \frac14} h_{\alpha\beta ,\rho} h_{\mu\nu ,\sigma} \!-\!
{\scriptstyle \frac14} h_{\alpha\rho ,\mu} h_{\beta\sigma ,\nu} \Bigr\} . \quad
\label{Linv}
\end{eqnarray}

We can read the graviton 3-point interaction off from expression (\ref{Linv}),
\begin{eqnarray}
\lefteqn{\mathcal{L}^{(3)} = {\scriptstyle (\frac{D}2 - 1)} \kappa H a^{D-1}
\Bigl\{{\scriptstyle \frac12} h h_{,\mu} h^{\mu 0} - h_{\alpha\beta} 
h^{\alpha\beta ,\mu} h_{\mu 0} - h^{\mu\nu} h_{,\mu} h_{\nu 0} \Bigr\} } 
\nonumber \\
& & \hspace{-.2cm} + \kappa a^{D-2} \Bigl\{{\scriptstyle \frac14} h h^{
\alpha\beta ,\mu} h_{\mu\alpha ,\beta} \!-\! h^{\alpha\beta} h_{\alpha\mu ,\nu}
h^{\mu\nu}_{~~,\beta} \!-\! {\scriptstyle \frac12} h^{\alpha\beta} h_{\alpha
\mu}^{~~,\nu} h_{\beta\nu}^{~~,\mu} \!-\! {\scriptstyle \frac14} h h_{,\mu} 
h^{\mu\nu}_{~~,\nu} \nonumber \\
& & + {\scriptstyle \frac12} h^{\alpha\beta} h_{\alpha\beta ,\mu} h^{\mu\nu}_{
~~,\nu} \!+\! {\scriptstyle \frac12} h^{\alpha\beta} h_{,\alpha} h_{\mu\beta
}^{~~,\mu} \!+\! {\scriptstyle \frac12} h^{\alpha\beta} h_{\alpha\mu ,\beta} 
h^{,\mu} \!+\! {\scriptstyle \frac18} h h^{,\mu} h_{,\mu} \!-\! {\scriptstyle 
\frac12} h^{\alpha\beta} h_{\alpha\beta ,\mu} h^{,\mu} \nonumber \\
& & \hspace{.2cm} - {\scriptstyle \frac14} h^{\alpha\beta} h_{,\alpha} 
h_{,\beta} \!-\! {\scriptstyle \frac18} h h^{\alpha\beta ,\mu} h_{\alpha\beta 
,\mu} \!+\! {\scriptstyle \frac12} h^{\alpha\beta} h_{\alpha\mu ,\nu} 
h_{\beta}^{~~\mu ,\nu} \!+\! {\scriptstyle \frac14} h^{\alpha\beta} h_{\mu\nu 
,\alpha} h^{\mu\nu}_{~~,\beta} \Bigl\} . \label{L3}
\end{eqnarray}
In deriving the associated vertex operators we must account for the 
indistinguishability of gravitons. This would ordinarily be accomplished by 
fully symmetrizing each interaction, which turns out to give over 70 distinct 
terms. For the pure graviton loop in Fig.~1 it is wasteful to first sum over 
these possibilities and then divide by the symmetry factor of 2 to compensate 
for overcounting. The more efficient strategy is to symmetrize the vertex
only on line \#1 and dispense with the symmetry factor. 

To obtain the partially symmetrized verticies one first takes any of the terms
from (\ref{L3}) and permutes graviton \#1 over the three possibilities. As an 
example, consider the term $\frac14 \kappa a^{D-2} h h^{\alpha\beta ,\mu}
h_{\mu\alpha , \beta}$. Denoting graviton \#1 by a breve, we obtain the 
following three terms,
\begin{equation}
\frac14 \kappa a^{D-2} {\breve h} h^{\alpha\beta ,\mu} h_{\mu\alpha ,\beta} 
\!+\! \frac14 \kappa a^{D-2} {\breve h_{\mu\alpha ,\beta}} h h^{\alpha\beta 
,\mu} \!+\! \frac14 \kappa a^{D-2} {\breve h^{\alpha\beta ,\mu}} h_{\mu\alpha 
,\beta} h \; . \label{sym}
\end{equation}
One then assigns the remaining two gravitons in each term as \#2 and \#3 in 
any way. For example, from (\ref{sym}) we could infer the following three 
vertex operators,
\begin{eqnarray}
\frac14 \kappa a^{D-2} \eta^{\alpha_1\beta_1} \partial_3^{(\alpha_2} 
\eta^{\beta_2)(\alpha_3} \partial_2^{\beta_3)} & , & \frac14 \kappa a^{D-2} 
\eta^{\alpha_2\beta_2} \partial_1^{(\alpha_3} \eta^{\beta_3)(\alpha_1} 
\partial_3^{\beta_1)} \nonumber \\
& {\rm and} & \frac14 \kappa a^{D-2} \eta^{\alpha_3\beta_3} \partial_2^{
(\alpha_1} \eta^{\beta_1)(\alpha_2} \partial_1^{\beta_2)} \; .
\end{eqnarray}
These are Vertex Operators \#10, \#11 and \#12, respectively, in Table 1.

\begin{table}

\vbox{\tabskip=0pt \offinterlineskip
\def\tablerule{\noalign{\hrule}}
\halign to460pt {\strut#& \vrule#\tabskip=1em plus2em& 
\hfil#& \vrule#& \hfil#\hfil& \vrule#& \hfil#& \vrule#& \hfil#\hfil& 
\vrule#\tabskip=0pt\cr
\tablerule
\omit&height4pt&\omit&&\omit&&\omit&&\omit&\cr
&&\omit\hidewidth \# &&\omit\hidewidth {\rm Vertex Operator}\hidewidth&& 
\omit\hidewidth \#\hidewidth&& \omit\hidewidth {\rm Vertex Operator}
\hidewidth&\cr
\omit&height4pt&\omit&&\omit&&\omit&&\omit&\cr
\tablerule
\omit&height2pt&\omit&&\omit&&\omit&&\omit&\cr
&& 1 && $\frac{(D-2)}4 \kappa H a^{D-1} \eta^{\alpha_1 \beta_1}  \eta^{\alpha_2 
\beta_2}  \partial_2^{(\alpha_3}  \delta_0^{\beta_3)}$ 
&& 22 && $\frac12 \kappa a^{D-2} \eta^{\alpha_2 (\alpha_3}  \eta^{\beta_3) 
\beta_2}  \partial_3^{(\alpha_1}  \partial_1^{\beta_1)}$ &\cr
\omit&height2pt&\omit&&\omit&&\omit&&\omit&\cr
\tablerule
\omit&height2pt&\omit&&\omit&&\omit&&\omit&\cr
&& 2 && $\frac{(D-2)}4 \kappa H a^{D-1} \eta^{\alpha_2 \beta_2}  \eta^{\alpha_3 
\beta_3}  \partial_3^{(\alpha_1}  \delta_0^{\beta_1)}$ 
&& 23 && $\frac12 \kappa a^{D-2} \eta^{\alpha_3 (\alpha_1}  \eta^{\beta_1) 
\beta_3}  \partial_1^{(\alpha_2}  \partial_2^{\beta_2)}$ &\cr
\omit&height2pt&\omit&&\omit&&\omit&&\omit&\cr
\tablerule
\omit&height2pt&\omit&&\omit&&\omit&&\omit&\cr
&& 3 && $\frac{(D-2)}4 \kappa H a^{D-1} \eta^{\alpha_3 \beta_3}  \eta^{\alpha_1 
\beta_1}  \partial_1^{(\alpha_2}  \delta_0^{\beta_2)}$ 
&& 24 && $\frac12 \kappa a^{D-2} \partial_2^{(\alpha_1}  \eta^{\beta_1) 
(\alpha_3}  \partial_3^{\beta_3)}  \eta^{\alpha_2 \beta_2}$ &\cr
\omit&height2pt&\omit&&\omit&&\omit&&\omit&\cr
\tablerule
\omit&height2pt&\omit&&\omit&&\omit&&\omit&\cr
&& 4 && $\!\!\! -\frac{(D-2)}2 \kappa H a^{D-1} \eta^{\alpha_1 (\alpha_2}  
\eta^{\beta_2) \beta_1} \partial_2^{(\alpha_3} \delta_0^{\beta_3)}\!\!\!\!$ 
&& 25 && $\frac12 \kappa a^{D-2} \partial_3^{(\alpha_2}  \eta^{\beta_2) 
(\alpha_1}  \partial_1^{\beta_1)}  \eta^{\alpha_3 \beta_3}$ &\cr
\omit&height2pt&\omit&&\omit&&\omit&&\omit&\cr
\tablerule
\omit&height2pt&\omit&&\omit&&\omit&&\omit&\cr
&& 5 && $\!\!\!-\frac{(D-2}2 \kappa H a^{D-1} \eta^{\alpha_2 (\alpha_3}  
\eta^{\beta_3) \beta_2} \partial_3^{(\alpha_1} \delta_0^{\beta_1)}\!\!\!\!$
&& 26 && $\frac12 \kappa a^{D-2} \partial_1^{(\alpha_3}  \eta^{\beta_3) 
(\alpha_2}  \partial_2^{\beta_2)}  \eta^{\alpha_1 \beta_1}$ &\cr
\omit&height2pt&\omit&&\omit&&\omit&&\omit&\cr
\tablerule
\omit&height2pt&\omit&&\omit&&\omit&&\omit&\cr
&& 6 && $\!\!\!-\frac{(D-2)}2 \kappa H a^{D-1} \eta^{\alpha_3 (\alpha_1}  
\eta^{\beta_1) \beta_3} \partial_1^{(\alpha_2} \delta_0^{\beta_2)}\!\!\!\!$
&& 27 && $\frac12 \kappa a^{D-2} \partial_2^{(\alpha_1}  \eta^{\beta_1) 
(\alpha_2}  \partial_3^{\beta_2)}  \eta^{\alpha_3 \beta_3}$ &\cr
\omit&height2pt&\omit&&\omit&&\omit&&\omit&\cr
\tablerule
\omit&height2pt&\omit&&\omit&&\omit&&\omit&\cr
&& 7 && $\!\!\!-\frac{(D-2)}2 \kappa H a^{D-1} \delta_0^{(\alpha_3} 
\eta^{\beta_3) (\alpha_1} \partial_2^{\beta_1)} \eta^{\alpha_2 \beta_2}
\!\!\!\!$ 
&& 28 && $\frac12 \kappa a^{D-2} \partial_3^{(\alpha_2}  \eta^{\beta_2) 
(\alpha_3}  \partial_1^{\beta_3)}  \eta^{\alpha_1 \beta_1}$ &\cr
\omit&height2pt&\omit&&\omit&&\omit&&\omit&\cr
\tablerule
\omit&height2pt&\omit&&\omit&&\omit&&\omit&\cr
&& 8 && $\!\!\!-\frac{(D-2)}2 \kappa H a^{D-1} \delta_0^{(\alpha_1} 
\eta^{\beta_1) (\alpha_2} \partial_3^{\beta_2)} \eta^{\alpha_3 \beta_3}
\!\!\!\!$ 
&& 29 && $\frac12 \kappa a^{D-2} \partial_1^{(\alpha_3}  \eta^{\beta_3) 
(\alpha_1}  \partial_2^{\beta_1)}  \eta^{\alpha_2 \beta_2}$ &\cr
\omit&height2pt&\omit&&\omit&&\omit&&\omit&\cr
\tablerule
\omit&height2pt&\omit&&\omit&&\omit&&\omit&\cr
&& 9 && $\!\!\!-\frac{(D-2)}2 \kappa H a^{D-1} \delta_0^{(\alpha_2} 
\eta^{\beta_2) (\alpha_3} \partial_1^{\beta_3)}  \eta^{\alpha_1 \beta_1}
\!\!\!\!$ 
&& 30 && $\frac18 \kappa a^{D-2} \eta^{\alpha_1 \beta_1}  \eta^{\alpha_2 
\beta_2}  \eta^{\alpha_3 \beta_3}  \partial_2 \cdot \partial_3$ &\cr
\omit&height2pt&\omit&&\omit&&\omit&&\omit&\cr
\tablerule
\omit&height2pt&\omit&&\omit&&\omit&&\omit&\cr
&& 10 && $\frac14 \kappa a^{D-2} \eta^{\alpha_1 \beta_1}  \partial_3^{
(\alpha_2}  \eta^{\beta_2) (\alpha_3}  \partial_2^{\beta_3)}$ 
&& 31 && $\frac14 \kappa a^{D-2} \eta^{\alpha_1 \beta_1}  \eta^{\alpha_2 
\beta_2}  \eta^{\alpha_3 \beta_3}  \partial_3 \cdot \partial_1$ &\cr
\omit&height2pt&\omit&&\omit&&\omit&&\omit&\cr
\tablerule
\omit&height2pt&\omit&&\omit&&\omit&&\omit&\cr
&& 11 && $\frac14 \kappa a^{D-2} \eta^{\alpha_2 \beta_2}  \partial_1^{
(\alpha_3}  \eta^{\beta_3) (\alpha_1}  \partial_3^{\beta_1)}$ 
&& 32 && $-\frac12 \kappa a^{D-2} \eta^{\alpha_1 (\alpha_2}  \eta^{\beta_2) 
\beta_1}  \eta^{\alpha_3 \beta_3}  \partial_2 \cdot \partial_3$ &\cr
\omit&height2pt&\omit&&\omit&&\omit&&\omit&\cr
\tablerule
\omit&height2pt&\omit&&\omit&&\omit&&\omit&\cr
&& 12 && $\frac14 \kappa a^{D-2} \eta^{\alpha_3 \beta_3}  \partial_2^{
(\alpha_1}  \eta^{\beta_1) (\alpha_2}  \partial_1^{\beta_2)}$ 
&& 33 && $-\frac12 \kappa a^{D-2} \eta^{\alpha_2 (\alpha_3}  \eta^{\beta_3) 
\beta_2}  \eta^{\alpha_1 \beta_1}  \partial_3 \cdot \partial_1$ &\cr
\omit&height2pt&\omit&&\omit&&\omit&&\omit&\cr
\tablerule
\omit&height2pt&\omit&&\omit&&\omit&&\omit&\cr
&& 13 && $-\kappa a^{D-2} \partial_3^{(\alpha_1}  \eta^{\beta_1) (\alpha_2} 
 \eta^{\beta_2) (\alpha_3}  \partial_2^{\beta_3)}$ 
&& 34 && $-\frac12 \kappa a^{D-2} \eta^{\alpha_3 (\alpha_1}  \eta^{\beta_1) 
\beta_3}  \eta^{\alpha_2 \beta_2}  \partial_1 \cdot \partial_2$ &\cr
\omit&height2pt&\omit&&\omit&&\omit&&\omit&\cr
\tablerule
\omit&height2pt&\omit&&\omit&&\omit&&\omit&\cr
&& 14 && $-\kappa a^{D-2} \partial_1^{(\alpha_2}  \eta^{\beta_2) (\alpha_3} 
 \eta^{\beta_3) (\alpha_1}  \partial_3^{\beta_1)}$ 
&& 35 && $-\frac14 \kappa a^{D-2} \partial_2^{(\alpha_1}  \partial_3^{
\beta_1)}  \eta^{\alpha_2 \beta_2}  \eta^{\alpha_3 \beta_3}$ &\cr
\omit&height2pt&\omit&&\omit&&\omit&&\omit&\cr
\tablerule
\omit&height2pt&\omit&&\omit&&\omit&&\omit&\cr
&& 15 && $-\kappa a^{D-2} \partial_2^{(\alpha_3}  \eta^{\beta_3) (\alpha_1} 
 \eta^{\beta_1) (\alpha_2}  \partial_1^{\beta_2)}$ 
&& 36 && $-\frac12 \kappa a^{D-2} \partial_3^{(\alpha_2}  \partial_1^{
\beta_2)}  \eta^{\alpha_3 \beta_3}  \eta^{\alpha_1 \beta_1}$ &\cr
\omit&height2pt&\omit&&\omit&&\omit&&\omit&\cr
\tablerule
\omit&height2pt&\omit&&\omit&&\omit&&\omit&\cr
&& 16 && $-\frac12 \kappa a^{D-2} \partial_3^{(\alpha_2}  \eta^{\beta_2) 
(\alpha_1}  \eta^{\beta_1) (\alpha_3}  \partial_2^{\beta_3)}$ 
&& 37 && $-\frac18 \kappa a^{D-2} \eta^{\alpha_1 \beta_1}  \eta^{\alpha_2 
(\alpha_3}  \eta^{\beta_3) \beta_2}  \partial_2 \cdot \partial_3$ &\cr
\omit&height2pt&\omit&&\omit&&\omit&&\omit&\cr
\tablerule
\omit&height2pt&\omit&&\omit&&\omit&&\omit&\cr
&& 17 && $-\kappa a^{D-2} \partial_1^{(\alpha_3} \eta^{\beta_3) (\alpha_2} 
 \eta^{\beta_2) (\alpha_1}  \partial_3^{\beta_1)}$ 
&& 38 && $-\frac14 \kappa a^{D-2} \eta^{\alpha_2 \beta_2} \eta^{\alpha_3 
(\alpha_1}  \eta^{\beta_1) \beta_3}  \partial_3 \cdot \partial_1$ &\cr
\omit&height2pt&\omit&&\omit&&\omit&&\omit&\cr
\tablerule
\omit&height2pt&\omit&&\omit&&\omit&&\omit&\cr
&& 18 && $-\frac14 \kappa a^{D-2} \eta^{\alpha_1\beta_1} \eta^{\alpha_2 
\beta_2}  \partial_2^{(\alpha_3}  \partial_3^{\beta_3)}$ 
&& 39 && $\frac12 \kappa a^{D-2} \eta^{\alpha_1) (\alpha_2} \eta^{\beta_2) 
(\alpha_3}  \eta^{\beta_3) (\beta_1}  \partial_2 \cdot \partial_3$ &\cr
\omit&height2pt&\omit&&\omit&&\omit&&\omit&\cr
\tablerule
\omit&height2pt&\omit&&\omit&&\omit&&\omit&\cr
&& 19 && $-\frac14 \kappa a^{D-2} \eta^{\alpha_2 \beta_2} \eta^{\alpha_3 
\beta_3}  \partial_3^{(\alpha_1}  \partial_1^{\beta_1)}$ 
&& 40 && $\kappa a^{D-2} \eta^{\alpha_1) (\alpha_2}  \eta^{\beta_2) 
(\alpha_3}  \eta^{\beta_3) (\beta_1}  \partial_3 \cdot \partial_1$ &\cr
\omit&height2pt&\omit&&\omit&&\omit&&\omit&\cr
\tablerule
\omit&height2pt&\omit&&\omit&&\omit&&\omit&\cr
&& 20 && $-\frac14 \kappa a^{D-2} \eta^{\alpha_3\beta_3} \eta^{\alpha_1 
\beta_1}  \partial_1^{(\alpha_2}  \partial_2^{\beta_2)}$ 
&& 41 && $\frac14 \kappa a^{D-2} \partial_2^{(\alpha_1} \partial_3^{
\beta_1)}  \eta^{\alpha_2 (\alpha_3}  \eta^{\beta_3) \beta_2}$ &\cr
\omit&height2pt&\omit&&\omit&&\omit&&\omit&\cr
\tablerule
\omit&height2pt&\omit&&\omit&&\omit&&\omit&\cr
&& 21 && $\frac12 \kappa a^{D-2} \eta^{\alpha_1 (\alpha_2} \eta^{\beta_2) 
\beta_1}  \partial_2^{(\alpha_3}  \partial_3^{\beta_3)}$
&& 42 && $\frac12 \kappa a^{D-2} \partial_3^{(\alpha_2}  \partial_1^{
\beta_2)}  \eta^{\alpha_3 (\alpha_1}  \eta^{\beta_1) \beta_3}$ &\cr 
\omit&height2pt&\omit&&\omit&&\omit&&\omit&\cr
\tablerule}}

\caption{Vertex operators contracted into $h_{\alpha_1\beta_1} h_{\alpha_2
\beta_2} h_{\alpha_3\beta_3}$ with $h_{\alpha_1\beta_1}$ external.}

\end{table}

Our gauge fixing term is an analogue of the de Donder term used in 
flat space \cite{TW3},
\begin{equation}
\mathcal{L}_{GF} = -\frac12 a^{D-2} \eta^{\mu\nu} F_{\mu} F_{\nu} \; , \;
F_{\mu} \equiv \eta^{\rho\sigma} \Bigl(h_{\mu\rho , \sigma} 
- \frac12 h_{\rho \sigma , \mu} + (D \!-\! 2) H a h_{\mu \rho}
\delta^0_{\sigma} \Bigr) . \label{GR}
\end{equation}
Because space and time components are treated differently it is useful to 
have an expression for the purely spatial parts of the Minkowski metric and
the Kronecker delta,
\begin{equation}
\overline{\eta}_{\mu\nu} \equiv \eta_{\mu\nu} + \delta^0_{\mu} \delta^0_{\nu}
\qquad {\rm and} \qquad \overline{\delta}^{\mu}_{\nu} \equiv \delta^{\mu}_{\nu}
- \delta_0^{\mu} \delta^0_{\nu} \; .
\end{equation}
The quadratic part of $\mathcal{L} + \mathcal{L}_{GF}$ can be partially
integrated to take the form $\frac12 h^{\mu\nu} D_{\mu\nu}^{~~\rho\sigma}
h_{\rho\sigma}$, where the kinetic operator is,
\begin{eqnarray}
\lefteqn{D_{\mu\nu}^{~~\rho\sigma} \equiv \left\{ \frac12 \overline{\delta}_{
\mu}^{~(\rho} \overline{\delta}_{\nu}^{~\sigma)} - \frac14 \eta_{\mu\nu} 
\eta^{\rho\sigma} - \frac1{2(D\!-\!3)} \delta_{\mu}^0 \delta_{\nu}^0
\delta_0^{\rho} \delta_0^{\sigma} \right\} D_A } \nonumber \\
& & \hspace{3cm} + \delta^0_{(\mu} \overline{\delta}_{\nu)}^{(\rho}
\delta_0^{\sigma)} \, D_B + \frac12 \Bigl(\frac{D\!-\!2}{D\!-\!3}\Bigr) 
\delta_{\mu}^0 \delta_{\nu}^0 \delta_0^{\rho} \delta_0^{\sigma} \, D_C 
\; , \qquad
\end{eqnarray}
and the three scalar differential operators are,
\begin{eqnarray}
D_A & \equiv & \partial_{\mu} \Bigl(\sqrt{-g} g^{\mu\nu} \partial_{\nu}\Bigr)
\; , \\
D_B & \equiv & \partial_{\mu} \Bigl(\sqrt{-g} g^{\mu\nu} \partial_{\nu}\Bigr)
- \frac1{D} \Bigl(\frac{D\!-\!2}{D\!-\!1}\Bigr) R \sqrt{-g} \; , \\
D_C & \equiv & \partial_{\mu} \Bigl(\sqrt{-g} g^{\mu\nu} \partial_{\nu}\Bigr)
- \frac2{D} \Bigl(\frac{D\!-\!3}{D\!-\!1}\Bigr) R \sqrt{-g} \; .
\end{eqnarray}
The associated ghost Lagrangian is,
\begin{eqnarray}
\lefteqn{\mathcal{L}_{\rm gh} \equiv - a^{D-2} \, \overline{\omega}^{\mu} 
{\delta F}_{\mu} \; ,} \\
& & = \overline{\omega}^{\mu} \Bigl( \overline{\delta}_{\mu}^{
~\nu} D_A \!+\! \delta^0_{\mu} \delta_0^{\nu} D_B\Bigr) \omega_{\nu} \!-\! 2 
\kappa a^{D-2} \overline{\omega}^{\mu , \nu} \Bigl( h^{\rho}_{~(\mu} \partial_{
\nu)} \!+\! {\scriptstyle \frac12} h_{\mu\nu}^{~~,\rho} \!-\! H a h_{\mu\nu}
\delta_0^{\rho}\Bigr) \omega_{\rho} \nonumber \\
& & \hspace{4cm} + \kappa \Bigl(a^{D-2} \overline{\omega}^{\mu}\Bigr)_{,\mu}
\Bigl(h^{\rho\sigma} \partial_{\sigma} \!+\! {\scriptstyle \frac12} h_{\sigma
}^{~\sigma , \rho} \!-\! H a h \delta_0^{\rho}\Bigr) \omega_{\rho} \; . \qquad
\label{ghost}
\end{eqnarray}

The ghost and graviton propagators in this gauge take the form of a sum
of constant index factors times scalar propagators,
\begin{eqnarray}
i\Bigl[{}_{\mu} \Delta_{\nu}\Bigr](x;x') & = & \overline{\eta}_{\mu\nu}
\, i\Delta_A(x;x') - \delta^0_{\mu} \delta^0_{\nu} \, i\Delta_B(x;x') 
\; , \label{ghprop} \\
i\Bigl[{}_{\mu\nu} \Delta_{\rho\sigma}\Bigr](x;x') & = & \sum_{I=A,B,C}
\Bigl[{}_{\mu\nu} T^I_{\rho\sigma}\Bigr] i\Delta_I(x;x') \; . \label{gprop}
\end{eqnarray}
The three scalar propagators invert the various scalar kinetic operators,
\begin{equation}
D_I \times i\Delta_I(x;x') = i \delta^D(x - x') \qquad {\rm for} \qquad
I = A,B,C \; , \label{sprops}
\end{equation}
and we will presently give explicit expressions for them. The index factors 
in the graviton propagator are,
\begin{eqnarray}
\Bigl[{}_{\mu\nu} T^A_{\rho\sigma}\Bigr] & = & 2 \, \overline{\eta}_{\mu (\rho}
\overline{\eta}_{\sigma) \nu} - \frac2{D\!-\! 3} \overline{\eta}_{\mu\nu}
\overline{\eta}_{\rho \sigma} \; , \\
\Bigl[{}_{\mu\nu} T^B_{\rho\sigma}\Bigr] & = & -4 \delta^0_{(\mu} 
\overline{\eta}_{\nu) (\rho} \delta^0_{\sigma)} \; , \\
\Bigl[{}_{\mu\nu} T^C_{\rho\sigma}\Bigr] & = & \frac2{(D \!-\!2) (D \!-\!3)}
\Bigl[(D \!-\!3) \delta^0_{\mu} \delta^0_{\nu} + \overline{\eta}_{\mu\nu}\Bigr]
\Bigl[(D \!-\!3) \delta^0_{\rho} \delta^0_{\sigma} + \overline{\eta}_{\rho
\sigma}\Bigr] \; .
\end{eqnarray}
With these definitions and equation (\ref{sprops}) for the scalar propagators
it is straightforward to verify that the graviton propagator (\ref{gprop})
indeed inverts the gauge-fixed kinetic operator,
\begin{equation}
D_{\mu\nu}^{~~\rho\sigma} \times i\Bigl[{}_{\rho\sigma} \Delta^{\alpha\beta}
\Bigr](x;x') = \delta_{\mu}^{(\alpha} \delta_{\nu}^{\beta)} i \delta^D(x-x')
\; .
\end{equation}

The scalar propagators can be expressed in terms of the following function 
of the invariant length $\ell(x;x')$ between $x^{\mu}$ and $x^{\prime \mu}$,
\begin{equation}
y(x;x') \equiv 4 \sin^2\Bigl(\frac12 H \ell(x;x')\Bigr) = a a' H^2 
\Bigl(\Vert \vec{x} \!-\! \vec{x}' \Vert^2 \!-\! (\vert\eta \!-\! \eta'
\vert - i \delta)^2\Bigr) \; .
\end{equation}
The most singular term for each case is the propagator for a massless,
conformally coupled scalar\cite{BD},
\begin{equation}
{i\Delta}_{\rm cf}(x;x') = \frac{H^{D-2}}{(4\pi)^{\frac{D}2}} \Gamma\Bigl(
\frac{D}2 \!-\! 1\Bigr) \Bigl(\frac4{y}\Bigr)^{\frac{D}2-1} \; .
\end{equation}
The $A$-type propagator obeys the same equation as that of a massless,
minimally coupled scalar. It has long been known that no de Sitter invariant
solution exists \cite{AF}. If one elects to break de Sitter invariance 
while preserving homogeneity and isotropy (this is known as the ``E(3)''
vacuum \cite{BA}), the minimal solution is \cite{OW1,OW2},
\begin{eqnarray}
\lefteqn{i \Delta_A(x;x') =  i \Delta_{\rm cf}(x;x') } \nonumber \\
& & + \frac{H^{D-2}}{(4\pi)^{\frac{D}2}} \frac{\Gamma(D \!-\! 1)}{\Gamma(
\frac{D}2)} \left\{\! \frac{D}{D\!-\! 4} \frac{\Gamma^2(\frac{D}2)}{\Gamma(D
\!-\! 1)} \Bigl(\frac4{y}\Bigr)^{\frac{D}2 -2} \!\!\!\!\!\! - \pi 
\cot\Bigl(\frac{\pi}2 D\Bigr) + \ln(a a') \!\right\} \nonumber \\
& & + \frac{H^{D-2}}{(4\pi)^{\frac{D}2}} \! \sum_{n=1}^{\infty}\! \left\{\!
\frac1{n} \frac{\Gamma(n \!+\! D \!-\! 1)}{\Gamma(n \!+\! \frac{D}2)} 
\Bigl(\frac{y}4 \Bigr)^n \!\!\!\! - \frac1{n \!-\! \frac{D}2 \!+\! 2} 
\frac{\Gamma(n \!+\!  \frac{D}2 \!+\! 1)}{\Gamma(n \!+\! 2)} \Bigl(\frac{y}4
\Bigr)^{n - \frac{D}2 +2} \!\right\} \! . \quad \label{DeltaA}
\end{eqnarray}
The B-type and $C$-type propagators possess de Sitter invariant (and also
unique) solutions \cite{RPW1},
\begin{eqnarray}
\lefteqn{i \Delta_B(x;x') =  i \Delta_{\rm cf}(x;x') - \frac{H^{D-2}}{(4
\pi)^{\frac{D}2}} \! \sum_{n=0}^{\infty}\! \left\{\!  \frac{\Gamma(n \!+\! D 
\!-\! 2)}{\Gamma(n \!+\! \frac{D}2)} \Bigl(\frac{y}4 \Bigr)^n \right. } 
\nonumber \\
& & \hspace{6.5cm} \left. - \frac{\Gamma(n \!+\!  \frac{D}2)}{\Gamma(n \!+\! 
2)} \Bigl( \frac{y}4 \Bigr)^{n - \frac{D}2 +2} \!\right\} \! , \qquad 
\label{DeltaB} \\
\lefteqn{i \Delta_C(x;x') =  i \Delta_{\rm cf}(x;x') + 
\frac{H^{D-2}}{(4\pi)^{\frac{D}2}} \! \sum_{n=0}^{\infty} \left\{\!
(n\!+\!1) \frac{\Gamma(n \!+\! D \!-\! 3)}{\Gamma(n \!+\! \frac{D}2)} 
\Bigl(\frac{y}4 \Bigr)^n \right. } \nonumber \\
& & \hspace{4.5cm} \left. - \Bigl(n \!-\! \frac{D}2 \!+\!  3\Bigr) \frac{
\Gamma(n \!+\! \frac{D}2 \!-\! 1)}{\Gamma(n \!+\! 2)} \Bigl(\frac{y}4 
\Bigr)^{n - \frac{D}2 +2} \!\right\} \! . \qquad \label{DeltaC}
\end{eqnarray}
These expressions might seem daunting but they are actually simple to use
because the infinite sums vanish in $D=4$, and each term in these sums
goes like a positive power of $y(x;x')$. This means the infinite
sums can only contribute when multiplied by a divergent term, and even
then only a small number of terms can contribute. Note also that the $B$-type
and $C$-type propagators agree with the conformal propagator in $D=4$.

The graviton-ghost-anti-ghost vertex operators can be read off from the 
order $\kappa$ terms of $\mathcal{L}_{\rm gh}$ in expression (\ref{ghost}).
Because the three fields are distinct there is no need for symmetrization.
Table 2 gives the ten vertex operators which result.

\begin{table}

\vbox{\tabskip=0pt \offinterlineskip
\def\tablerule{\noalign{\hrule}}
\halign to450pt {\strut#& \vrule#\tabskip=1em plus2em& 
\hfil#& \vrule#& \hfil#\hfil& \vrule#& \hfil#& \vrule#& \hfil#\hfil& 
\vrule#\tabskip=0pt\cr
\tablerule
\omit&height4pt&\omit&&\omit&&\omit&&\omit&\cr
&&\omit\hidewidth \# 
&&\omit\hidewidth {\rm Vertex Operator}\hidewidth&& 
\omit\hidewidth \#\hidewidth&& 
\omit\hidewidth {\rm Vertex Operator}
\hidewidth&\cr
\omit&height4pt&\omit&&\omit&&\omit&&\omit&\cr
\tablerule
\omit&height2pt&\omit&&\omit&&\omit&&\omit&\cr
&& 1 && $- \kappa a^{D-2} \eta^{\alpha_2 (\alpha_1} \eta^{\beta_1) 
\alpha_3} \partial_2 \cdot \partial_3$ 
&& 6 && $\frac12 \kappa a^{D-2} \eta^{\alpha_1 \beta_1} \partial_2^{
\alpha_2} \partial_1^{\alpha_3}$ &\cr
\omit&height2pt&\omit&&\omit&&\omit&&\omit&\cr
\tablerule
\omit&height2pt&\omit&&\omit&&\omit&&\omit&\cr
&& 2 && $- \kappa a^{D-2} \eta^{\alpha_3 (\alpha_1} \partial_2^{\beta_1)} 
\partial_3^{\alpha_2}$ 
&& 7 && $-\kappa H a^{D-1} \eta^{\alpha_1 \beta_1} \partial_2^{\alpha_2} 
\delta_0^{\alpha_3}$ &\cr
\omit&height2pt&\omit&&\omit&&\omit&&\omit&\cr
\tablerule
\omit&height2pt&\omit&&\omit&&\omit&&\omit&\cr
&& 3 && $- \kappa a^{D-2} \eta^{\alpha_2 (\alpha_1} \partial_2^{\beta_1)} 
\partial_1^{\alpha_3}$ 
&& 8 && $-{\scriptstyle (D-2)} \kappa H a^{D-1} \eta^{\alpha_3 (\alpha_1} 
\partial_3^{\beta_1)} \delta_0^{\alpha_2}$ &\cr
\omit&height2pt&\omit&&\omit&&\omit&&\omit&\cr
\tablerule
\omit&height2pt&\omit&&\omit&&\omit&&\omit&\cr
&& 4 && $2 \kappa H a^{D-1} \eta^{\alpha_2 (\alpha_1} \partial_2^{
\beta_1)} \delta_0^{\alpha_3}$ 
&& 9 && $-\frac{(D-2)}2 \kappa H a^{D-1} \eta^{\alpha_1 \beta_1} \partial_1^{
\alpha_3} \delta_0^{\alpha_2}$ &\cr
\omit&height2pt&\omit&&\omit&&\omit&&\omit&\cr
\tablerule
\omit&height2pt&\omit&&\omit&&\omit&&\omit&\cr
&& 5 && $\kappa a^{D-2} \eta^{\alpha_3 (\alpha_1} \partial_3^{\beta_1)} 
\partial_2^{\alpha_2}$ 
&& 10 && ${\scriptstyle (D-2)} \kappa H^2 a^D \eta^{\alpha_1 \beta_1} 
\delta_0^{\alpha_2} \delta_0^{\alpha_3}$ &\cr
\omit&height2pt&\omit&&\omit&&\omit&&\omit&\cr
\tablerule}}

\caption{Vertex operators contracted into $h_{\alpha_1\beta_1} 
{\overline \omega}_{\alpha_2} \omega_{\alpha_3}$.}

\end{table}

The final diagram in Fig.~1 represents a renormalization of the cosmological
constant. We compute it by expanding the relevant counterterm to first 
order in the graviton field,
\begin{equation}
-\frac{(D\!-\!2) \, \delta\Lambda}{16 \pi G} \sqrt{-g} = -\frac{(D\!-\!2) \,
\delta\Lambda a^D}{\kappa^2} \Bigl(1 + \frac12 \kappa h + \dots \Bigr) \; .
\end{equation}
Hence the final diagram of Fig.~1 makes the following contribution,
\begin{equation}
-\frac{i (\frac{D}2\!-\!1) \delta \Lambda a^D}{\kappa} \, \eta^{\alpha\beta} 
\; . \label{cosmo}
\end{equation}
By writing the sum of the first two diagrams in this form we can
express our final result as a graviton stress tensor for comparison 
with the computations of Ford \cite{LHF} and Finelli, Marozzi, Venturi and 
Vacca \cite{FMVV}.

\section{The Computation}

The purpose of this section is to describe the calculation. We begin by
explaining generally how one assembles the components of the previous
section to evaluate the first two diagrams of Fig.~1. We next give the
four contractions of each of the three index factors in the graviton
propagator. We also give the results of taking the coincidence limits
of zero, one and two derivatives of each of the three scalar propagators.
The graviton vertex operators turn out to possess a simple structure 
when organized into ten groups. A representative of each group is reduced.
Finally, the sum is taken and shown to give a small, positive shift in
the cosmological constant.

The graviton loop (first diagram of Fig.~1) consists of a sum of the 
coincidence limits of ($i$ times) the vertex operators from Table 1 acting
on the graviton propagator,
\begin{equation}
\Bigl({\rm Graviton\ Loop}\Bigr)^{\alpha\beta} = \sum_{i=1}^{42} 
\lim_{x' \rightarrow x} i V_i^{\alpha\beta\mu\nu\rho\sigma} \times 
i\Bigl[{}_{\mu\nu} \Delta_{\rho\sigma}\Bigr](x;x') \; .
\end{equation}
For example, the contribution from Vertex Operator \#1 in Table 1 is,
\begin{equation}
\Bigl({\rm Graviton\ Loop}\Bigr)^{\alpha\beta}_1 = \lim_{x' \rightarrow x}
i\Bigl(\frac{D\!-\!2}{4}\Bigr) \kappa H a^{D-1} \eta^{\alpha\beta} 
\eta^{\mu\nu} \partial^{\prime \rho} \delta_0^{\sigma} \times
i\Bigl[{}_{\mu\nu} \Delta_{\rho\sigma}\Bigr](x;x') \; .
\end{equation}
Similarly, the ghost loop (second diagram of Fig.~1) consists of minus the
sum of the coincidence limits of ($i$ times) the vertex operators from Table 
2 acting on the ghost propagator,
\begin{equation}
\Bigl({\rm Ghost\ Loop}\Bigr)^{\alpha\beta} = - \sum_{i=1}^{10} \lim_{x' 
\rightarrow x} i V_i^{\alpha\beta\mu\nu} \times i\Bigl[{}_{\mu} \Delta_{\nu}
\Bigr](x;x') \; .
\end{equation}
For example, the contribution from Vertex Operator \#1 on Table 2 is,
\begin{equation}
\Bigl({\rm Ghost\ Loop}\Bigr)^{\alpha\beta}_1 = \lim_{x' \rightarrow x}
i \kappa a^{D-2} \eta^{\alpha\mu} \eta^{\beta\nu} \partial \! \cdot \!
\partial' \times i\Bigl[{}_{\mu} \Delta_{\nu}\Bigr](x;x') \; .
\end{equation}
The only subtle point is that derivatives with respect to the external
line must be partially integrated back on the entire diagram. For example,
the contribution from Vertex Operator \#42 of Table 1 is,
\begin{equation}
\Bigl({\rm Graviton\ Loop}\Bigr)^{\alpha\beta}_{42} = -\partial^{\mu} \left\{
\lim_{x' \rightarrow x} \frac{i}2 \kappa a^{D-2} \eta^{\alpha\rho} 
\eta^{\beta\sigma} \partial^{\prime \nu} \times i\Bigl[{}_{\mu\nu} 
\Delta_{\rho\sigma}\Bigr](x;x') \right\} \; .
\end{equation}

From an examination of the vertex operators in Table 1 it is apparent that
we must take four generic contractions of the three index factors
$[\mbox{}_{\mu\nu} T^I_{\rho\sigma} ]$ which make up the graviton
propagator,
\begin{equation}
\eta^{\alpha\rho} \eta^{\beta\sigma} \eta^{\mu\nu} \quad , \quad
\eta^{\rho\sigma} \eta^{\mu\nu} \quad , \quad
\eta^{\mu\rho} \eta^{\nu\sigma} \quad , \quad
\eta^{\alpha\mu} \eta^{\nu\rho} \eta^{\sigma\beta} \; .
\end{equation}
For the $A$-type index factor these contractions give,
\begin{eqnarray}
\eta^{\alpha\rho} \eta^{\beta\sigma} \eta^{\mu\nu} \,
\Bigl[{}_{\mu\nu} T^A_{\sigma\rho}\Bigr] & = & - \frac{4}{D\!-\!3} \, 
\overline{\eta}^{\alpha\beta} \; , \nonumber \\
\eta^{\rho\sigma} \eta^{\mu\nu} \, \Bigl[{}_{\mu\nu} T^A_{\sigma\rho}\Bigr]
& = & -4 \Bigl(\frac{D\!-\!1}{D\!-\!3}\Bigr) \; , \nonumber \\
\eta^{\mu\rho} \eta^{\nu\sigma} \, \Bigl[{}_{\mu\nu} T^A_{\sigma\rho}\Bigr] 
& = & (D^2 \!-\! 3D \!-\! 2) \Bigl(\frac{D\!-\!1}{D\!-\!3}\Bigr) \; , 
\nonumber \\
\eta^{\alpha\mu} \eta^{\nu\rho} \eta^{\sigma\beta} \, \Bigl[{}_{\mu\nu} 
T^A_{\sigma\rho}\Bigr] & = & \Bigl(\frac{D^2 \!-\! 3D \!-\! 2}{D\!-\!3}\Bigr)
\, \overline{\eta}^{\alpha\beta} \; .
\end{eqnarray}
The four contractions of the $B$-type index factor are,
\begin{eqnarray}
\eta^{\alpha\rho} \eta^{\beta\sigma} \eta^{\mu\nu} \,
\Bigl[{}_{\mu\nu} T^B_{\sigma\rho}\Bigr] = 0 & , & \eta^{\rho\sigma} 
\eta^{\mu\nu} \, \Bigl[{}_{\mu\nu} T^B_{\sigma\rho}\Bigr] = 0 \; , \nonumber \\
\eta^{\mu\rho} \eta^{\nu\sigma} \, \Bigl[{}_{\mu\nu} T^B_{\sigma\rho}\Bigr] 
= 2 (D \!-\! 1) & , & \eta^{\alpha\mu} \eta^{\nu\rho} \eta^{\sigma\beta} \, 
\Bigl[{}_{\mu\nu} T^B_{\sigma\rho}\Bigr] = -(D\!-\!1) \delta_0^{\alpha} 
\delta_0^{\beta} + \overline{\eta}^{\alpha\beta} \; . \qquad 
\end{eqnarray}
And the four contractions of the $C$-type index factor give,
\begin{eqnarray}
\eta^{\alpha\rho} \eta^{\beta\sigma} \eta^{\mu\nu} \,
\Bigl[{}_{\mu\nu} T^C_{\sigma\rho}\Bigr] & = & \frac4{(D\!-\!2)(D\!-\!3)} \, 
\Bigl[(D\!-\!3) \delta_0^{\alpha} \delta_0^{\beta} + \overline{\eta}^{
\alpha\beta}\Bigr] \; , \nonumber \\
\eta^{\rho\sigma} \eta^{\mu\nu} \, \Bigl[{}_{\mu\nu} T^C_{\sigma\rho}\Bigr]
& = & \frac8{(D\!-\!2)(D\!-\!3)} \; , \nonumber \\
\eta^{\mu\rho} \eta^{\nu\sigma} \, \Bigl[{}_{\mu\nu} T^C_{\sigma\rho}\Bigr] 
& = & 2 \frac{(D^2 \!-\! 5D \!+\! 8)}{(D\!-\!2)(D\!-\!3} \; , \nonumber \\
\eta^{\alpha\mu} \eta^{\nu\rho} \eta^{\sigma\beta} \, \Bigl[{}_{\mu\nu} 
T^C_{\sigma\rho}\Bigr] & = & \frac2{(D\!-\!2)(D\!-\!3)} \, \Bigl[-(D\!-\!3)^2
\delta_0^{\alpha} \delta_0^{\beta} + \overline{\eta}^{\alpha\beta}\Bigr] \; .
\end{eqnarray}

We also require the coincidence limits of zero, one or two derivatives
acting on each of the scalar propagators. For the $A$-type propagator these
are,
\begin{eqnarray}
\lim_{x' \rightarrow x} \, {i\Delta}_A(x;x') & = & \frac{H^{D-2}}{(4\pi)^{
\frac{D}2}} \frac{\Gamma(D-1)}{\Gamma(\frac{D}2)} \left\{-\pi \cot\Bigl(
\frac{\pi}2 D \Bigr) + 2 \ln(a) \right\} , \\
\lim_{x' \rightarrow x} \, \partial_{\mu} {i\Delta}_A(x;x') & = & 
\frac{H^{D-2}}{(4\pi)^{\frac{D}2}} \frac{\Gamma(D-1)}{\Gamma(\frac{D}2)} 
\times H a \delta^0_{\mu} \; , \\
\lim_{x' \rightarrow x} \, \partial_{\mu} \partial_{\nu}' {i\Delta}_A(x;x') 
& = & \frac{H^{D-2}}{(4\pi)^{\frac{D}2}} \frac{\Gamma(D-1)}{\Gamma(\frac{D}2)} 
\times -\Bigl(\frac{D\!-\!1}{D}\Bigr) H^2 g_{\mu\nu} \; .
\end{eqnarray}
The analogous coincidence limits for the $B$-type propagator are actually 
finite in $D=4$ dimensions,
\begin{eqnarray}
\lim_{x' \rightarrow x} \, {i\Delta}_B(x;x') & = & \frac{H^{D-2}}{(4\pi)^{
\frac{D}2}} \frac{\Gamma(D-1)}{\Gamma(\frac{D}2)}\times -\frac1{D\!-\!2} \; ,\\
\lim_{x' \rightarrow x} \, \partial_{\mu} {i\Delta}_B(x;x') & = & 0 \; , \\
\lim_{x' \rightarrow x} \, \partial_{\mu} \partial_{\nu}' {i\Delta}_B(x;x') 
& = & \frac{H^{D-2}}{(4\pi)^{\frac{D}2}} \frac{\Gamma(D-1)}{\Gamma(\frac{D}2)} 
\times \frac1{D} H^2 g_{\mu\nu} \; .
\end{eqnarray}
The same is true for the coincidence limits of the $C$-type propagator,
\begin{eqnarray}
\lim_{x' \rightarrow x} \, {i\Delta}_C(x;x') & = & \frac{H^{D-2}}{(4\pi)^{
\frac{D}2}} \frac{\Gamma(D-1)}{\Gamma(\frac{D}2)}\times \frac1{(D\!-\!2)
(D\!-\!3)} \; ,\\
\lim_{x' \rightarrow x} \, \partial_{\mu} {i\Delta}_C(x;x') & = & 0 \; , \\
\lim_{x' \rightarrow x} \, \partial_{\mu} \partial_{\nu}' {i\Delta}_C(x;x') 
& = & \frac{H^{D-2}}{(4\pi)^{\frac{D}2}} \frac{\Gamma(D-1)}{\Gamma(\frac{D}2)} 
\times -\frac2{(D\!-\!2) D} H^2 g_{\mu\nu} \; . \qquad
\end{eqnarray}

Table 3 gives the contribution to the first diagram of Fig.~1 from each of the
42 graviton vertex operators. Although the 25 nonzero contributions might seem
bewilderingly varied they in fact derive from just ten distinct groups, each
of which sums to a simple result. We begin with Vertex Operators \#2 and \#5,
which derive from a single derivative of the $A$-type propagator. The reduction
of Vertex Operator \#2 is,
\begin{eqnarray}
\lefteqn{\Bigl({\rm Graviton\ Loop}\Bigr)^{\alpha\beta}_2 = i\Bigl(\frac{D\!-
\!2}{4}\Bigr) \kappa H a^{D-1} \eta^{\mu\nu} \eta^{\rho\sigma}\delta_0^{\alpha}
\partial^{\prime \beta} \times i\Bigl[{}_{\mu\nu} \Delta_{\rho\sigma}
\Bigr] \; , } \\
& & = i\Bigl(\frac{D\!- \!2}{4}\Bigr) \kappa H a^{D-1} \times -4 \Bigl(\frac{D
\!-\!1}{D\!-\!3}\Bigr) \times \frac{H^{D-2}}{(4\pi)^{\frac{D}2}} \frac{\Gamma(
D-1)}{\Gamma(\frac{D}2)} \times -H a \delta^{\alpha}_0 \delta^{\beta}_0 \; ,
\qquad \\
& & = \frac{i \kappa H^D a^D}{(4\pi)^{\frac{D}2}} \frac{\Gamma(D-1)}{\Gamma(
\frac{D}2)} \times \frac{(D\!-\!2) (D\!-\!1)}{(D\!-\!3)} \, \delta^{\alpha}_0 
\delta^{\beta}_0 \; .
\end{eqnarray}
The contribution from Vertex Operator \#5 is just $\frac12 (D^2\!-\!3D\!-\!2)$
times this, and the pole at $D=3$ cancels in their sum,
\begin{equation}
\Bigl({\rm Graviton\ Loop}\Bigr)^{\alpha\beta}_{2+5} = \frac{i \kappa H^D a^D}{
(4\pi)^{\frac{D}2}} \frac{\Gamma(D-1)}{\Gamma(\frac{D}2)} \times \frac12 
(D\!-\!2) (D\!-\!1) D \, \delta^{\alpha}_0 \delta^{\beta}_0 \; .
\end{equation}

\begin{table}

\vbox{\tabskip=0pt \offinterlineskip
\def\tablerule{\noalign{\hrule}}
\halign to460pt {\strut#& \vrule#\tabskip=1em plus2em& 
\hfil#& \vrule#& \hfil#\hfil& \vrule#& \hfil#& \vrule#& \hfil#\hfil& 
\vrule#\tabskip=0pt\cr
\tablerule
\omit&height4pt&\omit&&\omit&&\omit&&\omit&\cr
&&\omit\hidewidth \# &&\omit\hidewidth {\rm Vertex Contribution}\hidewidth&& 
\omit\hidewidth \#\hidewidth&& \omit\hidewidth {\rm Vertex Contribution}
\hidewidth&\cr
\omit&height4pt&\omit&&\omit&&\omit&&\omit&\cr
\tablerule
\omit&height2pt&\omit&&\omit&&\omit&&\omit&\cr
&& 1 && $0$
&& 22 && $-{\scriptstyle (D^2 - 3D - 2)} \frac{(D-1)^2}{2(D-3)} 
\delta^{\alpha}_0 \delta^{\beta}_0$ & \cr
\omit&height2pt&\omit&&\omit&&\omit&&\omit&\cr
\tablerule
\omit&height2pt&\omit&&\omit&&\omit&&\omit&\cr
&& 2 && $\frac{(D-2) (D-1)}{(D-3)} \delta^{\alpha}_0 \delta^{\beta}_0$
&& 23 && $0$ & \cr
\omit&height2pt&\omit&&\omit&&\omit&&\omit&\cr
\tablerule
\omit&height2pt&\omit&&\omit&&\omit&&\omit&\cr
&& 3 && $\frac{(D-1)}{(D-3)(D-2)} \eta^{\alpha\beta}$
&& 24 && $2 \frac{(D^2 -2D +2)}{(D-2)^2 D} \overline{\eta}^{\alpha\beta}
- \frac4{(D-2)^2 D} \delta^{\alpha}_0 \delta^{\beta}_0$ & \cr
\omit&height2pt&\omit&&\omit&&\omit&&\omit&\cr
\tablerule
\omit&height2pt&\omit&&\omit&&\omit&&\omit&\cr
&& 4 && $0$
&& 25 && $0$ & \cr
\omit&height2pt&\omit&&\omit&&\omit&&\omit&\cr
\tablerule
\omit&height2pt&\omit&&\omit&&\omit&&\omit&\cr
&& 5 && ${\scriptstyle (D^2 - 3D - 2)} \frac{(D-2) (D-1)}{2(D-3)} 
\delta^{\alpha}_0 \delta^{\beta}_0$
&& 26 && $0$ & \cr
\omit&height2pt&\omit&&\omit&&\omit&&\omit&\cr
\tablerule
\omit&height2pt&\omit&&\omit&&\omit&&\omit&\cr
&& 6 && $- \frac{(D - 1)}{(D-3)(D-2)} \overline{\eta}^{\alpha\beta}
- \frac{(D-1)}{(D-2)} \delta^{\alpha}_0 \delta^{\beta}_0$
&& 27 && $2 \frac{(D^2 -2D +2)}{(D-2)^2 D} \overline{\eta}^{\alpha\beta}
- \frac4{(D-2)^2 D} \delta^{\alpha}_0 \delta^{\beta}_0$ & \cr
\omit&height2pt&\omit&&\omit&&\omit&&\omit&\cr
\tablerule
\omit&height2pt&\omit&&\omit&&\omit&&\omit&\cr
&& 7 && $0$
&& 28 && $0$ & \cr
\omit&height2pt&\omit&&\omit&&\omit&&\omit&\cr
\tablerule
\omit&height2pt&\omit&&\omit&&\omit&&\omit&\cr
&& 8 && $0$
&& 29 && $0$ & \cr
\omit&height2pt&\omit&&\omit&&\omit&&\omit&\cr
\tablerule
\omit&height2pt&\omit&&\omit&&\omit&&\omit&\cr
&& 9 && $[-\frac12 {\scriptstyle (D-1)^2} + \frac{(D-1)}{(D-2)}] 
\eta^{\alpha \beta}$
&& 30 && $\frac{D}2 \frac{(D^2 - 3D + 4)}{(D-2)^2} \eta^{\alpha\beta}$ & \cr
\omit&height2pt&\omit&&\omit&&\omit&&\omit&\cr
\tablerule
\omit&height2pt&\omit&&\omit&&\omit&&\omit&\cr
&& 10 && ${\scriptstyle A(D)} \eta^{\alpha\beta}$
&& 31 && $-\frac{(D-1)^2}{(D-3)} \eta^{\alpha\beta}$ & \cr
\omit&height2pt&\omit&&\omit&&\omit&&\omit&\cr
\tablerule
\omit&height2pt&\omit&&\omit&&\omit&&\omit&\cr
&& 11 && $0$
&& 32 && $-2 \frac{(D^2 -2D +2)}{(D-2)^2} \overline{\eta}^{\alpha\beta}
+ \frac4{(D-2)^2} \delta^{\alpha}_0 \delta^{\beta}_0$ & \cr
\omit&height2pt&\omit&&\omit&&\omit&&\omit&\cr
\tablerule
\omit&height2pt&\omit&&\omit&&\omit&&\omit&\cr
&& 12 && $0$
&& 33 && $-{\scriptstyle (D^2 - 3D - 2)} \frac{(D-1)^2}{2(D-3)} 
\eta^{\alpha\beta}$ & \cr
\omit&height2pt&\omit&&\omit&&\omit&&\omit&\cr
\tablerule
\omit&height2pt&\omit&&\omit&&\omit&&\omit&\cr
&& 13 && ${\scriptstyle B(D)} \overline{\eta}^{\alpha\beta} + {\scriptstyle
C(D)} \delta^{\alpha}_0 \delta^{\beta}_0$
&& 34 && $2 \frac{(D-1)}{(D-3)} \overline{\eta}^{\alpha\beta}$ & \cr
\omit&height2pt&\omit&&\omit&&\omit&&\omit&\cr
\tablerule
\omit&height2pt&\omit&&\omit&&\omit&&\omit&\cr
&& 14 && $0$
&& 35 && $- \frac{(D^2 - 3D + 4)}{(D-2)^2} \eta^{\alpha\beta}$ & \cr
\omit&height2pt&\omit&&\omit&&\omit&&\omit&\cr
\omit&height2pt&\omit&&\omit&&\omit&&\omit&\cr
\tablerule
\omit&height2pt&\omit&&\omit&&\omit&&\omit&\cr
&& 15 && $0$
&& 36 && $0$ & \cr
\omit&height2pt&\omit&&\omit&&\omit&&\omit&\cr
\tablerule
\omit&height2pt&\omit&&\omit&&\omit&&\omit&\cr
&& 16 && $\frac12 {\scriptstyle B(D)} \overline{\eta}^{\alpha\beta} + 
\frac12 {\scriptstyle C(D)} \delta^{\alpha}_0 \delta^{\beta}_0$
&& 37 && $-\frac{D}2 {\scriptstyle A(D)} \eta^{\alpha\beta}$ & \cr
\omit&height2pt&\omit&&\omit&&\omit&&\omit&\cr
\tablerule
\omit&height2pt&\omit&&\omit&&\omit&&\omit&\cr
&& 17 && $0$
&& 38 && $\frac{(D-1)}{(D-3)} \overline{\eta}^{\alpha\beta}$ & \cr
\omit&height2pt&\omit&&\omit&&\omit&&\omit&\cr
\tablerule
\omit&height2pt&\omit&&\omit&&\omit&&\omit&\cr
&& 18 && $- \frac{(D^2 - 3D + 4)}{(D-2)^2} \eta^{\alpha\beta}$
&& 39 && $-\frac{D}2 {\scriptstyle B(D)} \overline{\eta}^{\alpha\beta} -
\frac{D}2 {\scriptstyle C(D)} \delta^{\alpha}_0 \delta^{\beta}_0$ & \cr
\omit&height2pt&\omit&&\omit&&\omit&&\omit&\cr
\tablerule
\omit&height2pt&\omit&&\omit&&\omit&&\omit&\cr
&& 19 && $-\frac{(D-1)^2}{(D-3)} \delta^{\alpha}_0 \delta^{\beta}_0$
&& 40 && ${\scriptstyle (D^2 -3D -2)} \frac{(D-1)}{(D-3)} \overline{\eta}^{
\alpha\beta}$ & \cr
\omit&height2pt&\omit&&\omit&&\omit&&\omit&\cr
\tablerule
\omit&height2pt&\omit&&\omit&&\omit&&\omit&\cr
&& 20 && $0$
&& 41 && ${\scriptstyle A(D)} \eta^{\alpha\beta}$ & \cr
\omit&height2pt&\omit&&\omit&&\omit&&\omit&\cr
\tablerule
\omit&height2pt&\omit&&\omit&&\omit&&\omit&\cr
&& 21 && $2 \frac{(D^2 -2D +2)}{(D-2)^2 D} \overline{\eta}^{\alpha\beta}
- \frac4{(D-2)^2 D} \delta^{\alpha}_0 \delta^{\beta}_0$
&& 42 && $0$ & \cr
\omit&height2pt&\omit&&\omit&&\omit&&\omit&\cr
\tablerule}}

\caption{Contributions from the graviton loop with an overall factor of 
$\frac{i \kappa H^D a^D}{(4\pi)^{\frac{D}2}} \frac{\Gamma(D-1)}{\Gamma(\frac{D
}2)}$ removed. The three constants are ${\scriptstyle A(D) \equiv - \frac14 
(D-1)^2 + 1 + \frac1{(D-2)^2}}$, ${\scriptstyle B(D) \equiv (D-1) - \frac1{D} 
- \frac1{(D-2)} - \frac2{(D-2)^2}}$ and ${\scriptstyle C(D) \equiv 1 +
\frac2{D} - \frac3{(D-2)} + \frac2{(D-2)^2}}$.}

\end{table}

The contributions from Vertex Operators \#19 and \#22 involve a partially 
integrated derivative acting back on a derivative of the $A$-type propagator. 
The contribution from Vertex Operator \#19 is,
\begin{eqnarray}
\lefteqn{\Bigl({\rm Graviton\ Loop}\Bigr)^{\alpha\beta}_{19} = -\partial^{
\alpha} \left\{ -\frac{i}4 \kappa a^{D-2} \eta^{\mu\nu} \eta^{\rho\sigma}
\partial^{\prime \beta} \times i\Bigl[{}_{\mu\nu} \Delta_{\rho\sigma}
\Bigr]\right\} \; , } \\
& & = \partial_0 \left\{ -\frac{i}4 \kappa H a^{D-2} \times -4 \Bigl(\frac{D
\!-\!1}{D\!-\!3}\Bigr) \times \frac{H^{D-2}}{(4\pi)^{\frac{D}2}} \frac{\Gamma(
D-1)}{\Gamma(\frac{D}2)} \times -H a \delta^{\alpha}_0 \delta^{\beta}_0 
\right\} , \qquad \\
& & = \frac{i \kappa H^D a^D}{(4\pi)^{\frac{D}2}} \frac{\Gamma(D-1)}{\Gamma(
\frac{D}2)} \times - \frac{(D\!-\!1)^2}{(D\!-\!3)} \, \delta^{\alpha}_0 
\delta^{\beta}_0 \; .
\end{eqnarray}
The contribution from Vertex Operator \#22 is just $\frac12 (D^2\!-\!3D\!-\!2)$
times this, and the pole at $D=3$ again cancels in their sum,
\begin{equation}
\Bigl({\rm Graviton\ Loop}\Bigr)^{\alpha\beta}_{19+22} = \frac{i \kappa H^D a^D
}{(4\pi)^{\frac{D}2}} \frac{\Gamma(D-1)}{\Gamma(\frac{D}2)} \times -\frac12 
(D\!-\!1)^2 D \, \delta^{\alpha}_0 \delta^{\beta}_0 \; .
\end{equation}
Hence the first four vertex operators we have considered contribute,
\begin{equation}
\Bigl({\rm Graviton\ Loop}\Bigr)^{\alpha\beta}_{{2+5} \atop {+19+22}} = 
\frac{i \kappa H^D a^D}{(4\pi)^{\frac{D}2}} \frac{\Gamma(D-1)}{\Gamma(
\frac{D}2)} \times -\frac12 (D\!-\!1) D \, \delta^{\alpha}_0 \delta^{\beta}_0 
\; .
\end{equation}

The contributions from Vertex Operators \#31 and \#33 are proportional to
$\eta^{\alpha\beta}$ with a partially integrated derivative contracted into
a derivative of the $A$-type propagator. The contribution from Vertex Operator 
\#31 is,
\begin{eqnarray}
\lefteqn{\Bigl({\rm Graviton\ Loop}\Bigr)^{\alpha\beta}_{31} = -\partial_{
\gamma} \left\{\frac{i}4 \kappa a^{D-2} \eta^{\alpha\beta} \eta^{\mu\nu} 
\eta^{\rho\sigma} \partial^{\prime \gamma} \times i\Bigl[{}_{\mu\nu} 
\Delta_{\rho\sigma} \Bigr]\right\} \; , } \\
& & = -\partial_0 \left\{\frac{i}4 \kappa H a^{D-2} \times -4 \Bigl(\frac{D
\!-\!1}{D\!-\!3}\Bigr) \times \frac{H^{D-2}}{(4\pi)^{\frac{D}2}} \frac{\Gamma(
D-1)}{\Gamma(\frac{D}2)} \times -H a \eta^{\alpha\beta} \right\} , \qquad \\
& & = \frac{i \kappa H^D a^D}{(4\pi)^{\frac{D}2}} \frac{\Gamma(D-1)}{\Gamma(
\frac{D}2)} \times - \frac{(D\!-\!1)^2}{(D\!-\!3)} \, \eta^{\alpha\beta} \; .
\end{eqnarray}
In what must by now seem a familiar pattern, the contribution from Vertex 
Operator \#33 is just $\frac12 (D^2\!-\!3D\!-\!2)$ times this, and the pole 
at $D=3$ cancels in their sum,
\begin{equation}
\Bigl({\rm Graviton\ Loop}\Bigr)^{\alpha\beta}_{31+33} = \frac{i \kappa H^D a^D
}{(4\pi)^{\frac{D}2}} \frac{\Gamma(D-1)}{\Gamma(\frac{D}2)} \times -\frac12 
(D\!-\!1)^2 D \, \eta^{\alpha\beta} \; .
\end{equation}

The contributions from Vertex Operators \#34, \#38 and \#40 also have a 
partially integrated derivative contracted into a derivative of the $A$-type 
propagator, but their free indicies reside inside the propagator. The 
contribution from Vertex Operator \#34 is,
\begin{eqnarray}
\lefteqn{\Bigl({\rm Graviton\ Loop}\Bigr)^{\alpha\beta}_{34} = -\partial_{
\gamma} \left\{-\frac{i}2 \kappa a^{D-2} \eta^{\alpha\rho} \eta^{\mu\nu} 
\eta^{\beta\sigma} \partial^{\prime \gamma} \times i\Bigl[{}_{\mu\nu} 
\Delta_{\rho\sigma} \Bigr]\right\} \; , } \\
& & = -\partial_0 \left\{-\frac{i}2 \kappa H a^{D-2} \times \Bigl(\frac{-4}{
D\!-\!3}\Bigr) \, \overline{\eta}^{\alpha\beta} \times \frac{H^{D-2}}{(4\pi)^{
\frac{D}2}} \frac{\Gamma(D-1)}{\Gamma(\frac{D}2)} \times -H a \right\} , 
\qquad \\
& & = \frac{i \kappa H^D a^D}{(4\pi)^{\frac{D}2}} \frac{\Gamma(D-1)}{\Gamma(
\frac{D}2)} \times 2 \Bigl(\frac{D\!-\!1}{D\!-\!3}\Bigr) \, \overline{\eta}^{
\alpha\beta} \; .
\end{eqnarray}
The contribution from Vertex Operator \#38 is half this, and the contribution
from \#40 is $\frac12 (D^2\!-\!3D\!-\!2)$ times it. Hence the three
contributions sum to,
\begin{equation}
\Bigl({\rm Graviton\ Loop}\Bigr)^{\alpha\beta}_{34+38+40} = \frac{i \kappa H^D 
a^D }{(4\pi)^{\frac{D}2}} \frac{\Gamma(D-1)}{\Gamma(\frac{D}2)} \left\{
(D\!-\!1) D \!+\! \Bigl(\frac{D\!-\!1}{D\!-\!3}\Bigr) \right\} 
\overline{\eta}^{\alpha\beta} .
\end{equation}

Vertex Operator \ is one of those with only a single derivative. It 
contributes,
\begin{eqnarray}
\lefteqn{\Bigl({\rm Graviton\ Loop}\Bigr)^{\alpha\beta}_6 = -\delta_0^{\mu}
\partial^{\nu} \left\{-i\Bigl(\frac{D\!-\!2}2\Bigr) \kappa a^{D-1} 
\eta^{\alpha\rho} \eta^{\beta\sigma} \times i\Bigl[{}_{\mu\nu} \Delta_{\rho
\sigma} \Bigr]\right\} \; , } \\
& & \hspace{-.5cm} = \partial_0 \! \left\{-i\Bigl(\frac{D\!-\!2}2\Bigr) \kappa 
a^{D-1} \! \times \!  \Bigl[\mbox{}_{00} T_C^{\alpha\beta}\Bigr] \! \times \!
\frac{H^{D-2}}{(4\pi)^{\frac{D}2}} \frac{\Gamma(D-1)}{\Gamma(\frac{D}2)} \!
\times \! \frac1{(D\!-\!3) (D\!-\!2)} \right\} , \qquad \\
& & \hspace{-.5cm} = \frac{i \kappa H^D a^D}{(4\pi)^{\frac{D}2}} \frac{\Gamma(
D-1)}{\Gamma(\frac{D}2)} \left\{ -\frac{(D\!-\!1)}{(D\!-\!3)(D\!-\!2 )}\, 
\overline{\eta}^{\alpha\beta} -\Bigl(\frac{D\!-\!1}{D\!-\!2}\Bigr) \,
\delta_0^{\alpha} \delta_0^{\beta} \right\} .
\end{eqnarray}
The sum of this with the three previous terms is free of the pole at $D=3$,
\begin{equation}
\Bigl({\rm Graviton\ Loop}\Bigr)^{\alpha\beta}_{{6+34} \atop {+38+40}} = 
\frac{i \kappa H^D a^D }{(4\pi)^{\frac{D}2}} \frac{\Gamma(D-1)}{\Gamma(\frac{
D}2)} \left\{(D\!-\!1) D \overline{\eta}^{\alpha\beta} + \Bigl(\frac{D\!-\!1}{
D\!-\!2}\Bigr) \eta^{\alpha\beta} \right\} .
\end{equation}

The contributions from Vertex Operators \#10, \#37 and \#41 are all 
proportional to $\eta^{\alpha \beta}$ times the coincidence limit of a double 
derivative. The reduction for \#10 is,
\begin{eqnarray}
\lefteqn{\Bigl({\rm Graviton\ Loop}\Bigr)^{\alpha\beta}_{10} = \frac{i}4
\kappa a^{D-2} \eta^{\alpha\beta} \eta^{\mu\rho} \partial^{\sigma} 
\partial^{\prime \nu} \times i\Bigl[{}_{\mu\nu} \Delta_{\rho \sigma} \Bigr]
\; , } \\
& & = \frac{i}4 \kappa a^{D-2} \eta^{\alpha\beta} \times \frac{H^{D-2}}{(4\pi
)^{\frac{D}2}} \frac{\Gamma(D-1)}{\Gamma(\frac{D}2)} \times H^2 a^2 \times
\eta^{\mu\rho} \eta^{\nu\sigma} \nonumber \\
& & \hspace{2cm} \times \left\{ -\Bigl(\frac{D\!-\!1}{D}\Bigr) \Bigl[{}_{\mu
\nu} T^A_{\rho \sigma}\Bigr] + \frac1{D} \Bigl[{}_{\mu\nu} T^B_{\rho \sigma}
\Bigr] - \frac2{(D\!-\!2)D} \Bigl[{}_{\mu\nu} T^C_{\rho \sigma}\Bigr]
\right\} , \qquad \\
& & = \frac{i \kappa H^D a^D}{(4\pi)^{\frac{D}2}} \frac{\Gamma(D-1)}{\Gamma(
\frac{D}2)} \left\{ -\frac14 (D\!-\!1)^2 + 1 + \frac1{(D\!-\!2)^2} \right\}
\eta^{\alpha\beta} \; . \label{10}
\end{eqnarray}
We define the bracketed constant in the final expression as $A(D)$. Vertex
Operator \#41 gives the same, and \#37 gives $-\frac{D}2$ times that of \#10.
So the three of them sum to be $-\frac12(D\!-\!4)$ times (\ref{10}).

The contributions from Vertex Operators \#18, \#30 and \#35 are also 
proportional to $\eta^{\alpha \beta}$ times the coincidence limit of a double 
derivative, but with the other tensor contraction of the propagator. The 
reduction for \#18 is,
\begin{eqnarray}
\lefteqn{\Bigl({\rm Graviton\ Loop}\Bigr)^{\alpha\beta}_{18} = -\frac{i}4
\kappa a^{D-2} \eta^{\alpha\beta} \eta^{\mu\nu} \partial^{\rho} 
\partial^{\prime \sigma} \times i\Bigl[{}_{\mu\nu} \Delta_{\rho \sigma} \Bigr]
\; , } \\
& & = -\frac{i}4 \kappa a^{D-2} \eta^{\alpha\beta} \times \frac{H^{D-2}}{(4\pi
)^{\frac{D}2}} \frac{\Gamma(D-1)}{\Gamma(\frac{D}2)} \times H^2 a^2 \times
\eta^{\mu\nu} \eta^{\rho\sigma} \nonumber \\
& & \hspace{2cm} \times \left\{ -\Bigl(\frac{D\!-\!1}{D}\Bigr) \Bigl[{}_{\mu
\nu} T^A_{\rho \sigma}\Bigr] + \frac1{D} \Bigl[{}_{\mu\nu} T^B_{\rho \sigma}
\Bigr] - \frac2{(D\!-\!2)D} \Bigl[{}_{\mu\nu} T^C_{\rho \sigma}\Bigr]
\right\} , \qquad \\
& & = \frac{i \kappa H^D a^D}{(4\pi)^{\frac{D}2}} \frac{\Gamma(D-1)}{\Gamma(
\frac{D}2)} \left\{-1 - \frac1{(D\!-\!2)} - \frac2{(D\!-\!2)^2} \right\}
\eta^{\alpha\beta} \; . \label{18}
\end{eqnarray}
Vertex Operator \#35 gives the same, and \#30 gives $-\frac{D}2$, so the three
of them again sum to be $-\frac12 (D\!-\!4)$ times the first. Of course we
can add these to the preceding three to find,
\begin{eqnarray}
\lefteqn{\Bigl({\rm Graviton\ Loop}\Bigr)^{\alpha\beta}_{{10+37+41} \atop 
{+18+30+35}} = \frac{i \kappa H^D a^D }{(4\pi)^{\frac{D}2}} \frac{\Gamma(D-1)}{
\Gamma(\frac{D}2)} } \nonumber \\
& & \hspace{2cm} \times \frac12 (D\!-\!4) \left\{\frac14 (D\!-\!1)^2 + 
\frac1{(D\!- \!2)} + \frac1{(D\!-\!2)^2} \right\} \eta^{\alpha\beta} . \qquad 
\label{D-4}
\end{eqnarray}

The contributions from Vertex Operators \#13, \#16 and \#39 involve the 
coincidence limit of a double derivative, but times one of the contractions in 
which free indicies reside on the propagator. The reduction for \#13 is,
\begin{eqnarray}
\lefteqn{\Bigl({\rm Graviton\ Loop}\Bigr)^{\alpha\beta}_{13} = -i \kappa 
a^{D-2} \eta^{\alpha\mu} \eta^{\nu\rho} \partial^{\sigma} \partial^{\prime 
\beta} \times i\Bigl[{}_{\mu\nu} \Delta_{\rho \sigma} \Bigr] \; , } \\
& & = -i \kappa a^{D-2} \times \frac{H^{D-2}}{(4\pi)^{\frac{D}2}} \frac{
\Gamma(D-1)}{\Gamma(\frac{D}2)} \times H^2 a^2 \times \eta^{\alpha \mu} 
\eta^{\nu\rho} \eta^{\beta\sigma} \nonumber \\
& & \hspace{2cm} \times \left\{ -\Bigl(\frac{D\!-\!1}{D}\Bigr) \Bigl[{}_{\mu
\nu} T^A_{\rho \sigma}\Bigr] + \frac1{D} \Bigl[{}_{\mu\nu} T^B_{\rho \sigma}
\Bigr] - \frac2{(D\!-\!2)D} \Bigl[{}_{\mu\nu} T^C_{\rho \sigma}\Bigr]
\right\} , \qquad \\
& & = \frac{i \kappa H^D a^D}{(4\pi)^{\frac{D}2}} \frac{\Gamma(D-1)}{\Gamma(
\frac{D}2)} \left\{B(D) \, \overline{\eta}^{\alpha\beta} + C(D) \, \delta_0^{
\alpha} \delta_0^{\beta} \right\} , \label{13}
\end{eqnarray}
where the $D$-dependent constants in (\ref{13}) are,
\begin{eqnarray}
B(D) & = & (D\!-\!1) - \frac2{D} - \frac1{(D\!-\!2)} - \frac2{(D\!-\!2)^2} 
\; , \\
C(D) & = & 1 + \frac2{D} - \frac3{(D\!-\!2)} + \frac2{(D\!-\!2)^2} \; .
\end{eqnarray}
Vertex Operator \#16 gives half of (\ref{13}), and \#39 gives $-\frac{D}2$ 
times (\ref{13}), so the three contributions sum to $-\frac12 (D\!-\!3)$
times (\ref{13}).

The contributions from Vertex Operators \#21, \#24, \#27 and \#32 have the
same structure but with the other contraction of the propagator. The reduction 
for \#21 is,
\begin{eqnarray}
\lefteqn{\Bigl({\rm Graviton\ Loop}\Bigr)^{\alpha\beta}_{21} = \frac{i}2 \kappa 
a^{D-2} \eta^{\alpha\mu} \eta^{\beta\nu} \partial^{\rho} \partial^{\prime 
\sigma} \times i\Bigl[{}_{\mu\nu} \Delta_{\rho \sigma} \Bigr] \; , } \\
& & = \frac{i}2 \kappa a^{D-2} \times \frac{H^{D-2}}{(4\pi)^{\frac{D}2}} \frac{
\Gamma(D-1)}{\Gamma(\frac{D}2)} \times H^2 a^2 \times \eta^{\alpha \mu} 
\eta^{\beta\nu} \eta^{\rho\sigma} \nonumber \\
& & \hspace{2cm} \times \left\{ -\Bigl(\frac{D\!-\!1}{D}\Bigr) \Bigl[{}_{\mu
\nu} T^A_{\rho \sigma}\Bigr] + \frac1{D} \Bigl[{}_{\mu\nu} T^B_{\rho \sigma}
\Bigr] - \frac2{(D\!-\!2)D} \Bigl[{}_{\mu\nu} T^C_{\rho \sigma}\Bigr]
\right\} , \qquad \\
& & = \frac{i \kappa H^D a^D}{(4\pi)^{\frac{D}2}} \frac{\Gamma(D-1)}{\Gamma(
\frac{D}2)} \left\{2 \frac{(D^2\!-\!2D\!+\!2)}{(D\!-\!2)^2 D} \, 
\overline{\eta}^{\alpha\beta} - \frac4{(D\!-\!2)^2 D} \, \delta_0^{\alpha} 
\delta_0^{\beta} \right\} , \qquad \label{21}
\end{eqnarray}
Vertex Operators \#24 and \#27 each give the same, and \#32 gives $-D$ times
(\ref{21}), so the four contributions sum to $-(D\!-\!3)$ times (\ref{21}).
At this stage we note that the pole at $D\!=\!0$ cancels when the contributions
from the preceding seven vertex operators are summed,
\begin{eqnarray}
\lefteqn{\Bigl({\rm Graviton\ Loop}\Bigr)^{\alpha\beta}_{{13+16+39} \atop 
{+21+24+27+32}} = \frac{i \kappa H^D a^D }{(4\pi)^{\frac{D}2}} \frac{
\Gamma(D-1)}{\Gamma(\frac{D}2)} } \nonumber \\
& & \hspace{.5cm} \times \left\{-\frac12 (D\!-\!3) D \, \overline{\eta}^{
\alpha\beta} + \frac12 (D\!-\!3) \Bigl[1 - \frac1{(D\!- \!2)} - \frac2{(D\!-\!
2)^2} \Bigr] \eta^{\alpha\beta} \right\} . \qquad 
\end{eqnarray}
Adding this to (\ref{D-4}) results in cancellation of the double pole at
$D\!=\!2$,
\begin{eqnarray}
\lefteqn{\Bigl({\rm Graviton\ Loop}\Bigr)^{\alpha\beta}_{{10+37+41} \atop 
{+18+30+35}} \!\!\!+ \Bigl({\rm Graviton\ Loop}\Bigr)^{\alpha\beta}_{{13+16+39}
\atop {+21+24+27+32}} \!\!\!\! = \frac{i \kappa H^D a^D }{(4\pi)^{\frac{D}2}} 
\frac{\Gamma(D-1)}{\Gamma(\frac{D}2)} } \nonumber \\
& & \times \left\{ -\frac12 (D\!-\!3) D \, \overline{\eta}^{\eta\beta} +
\Bigl[-\frac1{(D\!-\!2)} \!+\! \frac{D}2 \!-\! \frac32 \!+\! \frac18 (D\!-\!4)
(D\!-\!1)^2 \Bigr] \eta^{\eta\beta} \right\} ! \qquad
\end{eqnarray}

Finally, Vertex Operators \#3 and \#9 are proportional to $\eta^{\alpha\beta}$
times a single derivative integrated back on the coincidence limit of the
undifferentiated propagator. For Vertex Operator \#3 the reduction is,
\begin{eqnarray}
\lefteqn{\Bigl({\rm Graviton\ Loop}\Bigr)^{\alpha\beta}_3 = -\delta_0^{\mu}
\partial^{\nu} \left\{i \Bigl(\frac{D\!-\!2}4\Bigr) \kappa H a^{D-1} \eta^{
\alpha\beta} \eta^{\rho\sigma} \times i\Bigl[{}_{\mu\nu} \Delta_{\rho \sigma} 
\Bigr] \right\} \; , } \\
& & \hspace{-.5cm} = \partial_0 \left\{i \Bigl(\frac{D\!-\!2}4\Bigr) \kappa H 
a^{D-1} \!\! \times \! \frac{4 \eta^{\alpha\beta}}{(D\!-\!2)} \! \times \!
\frac{H^{D-2}}{(4\pi)^{\frac{D}2}} \frac{\Gamma(D-1)}{\Gamma(\frac{D}2)} 
\! \times \! \frac1{(D\!-\!3)(D\!-\!2)} \right\} \! , \quad \\
& & \hspace{-.5cm} = \frac{i \kappa H^D a^D }{(4\pi)^{\frac{D}2}} 
\frac{\Gamma(D-1)}{\Gamma(\frac{D}2)} \times \frac{(D\!-\!1)}{(D\!-\!3)
(D\!-\!2)} \, \eta^{\alpha\beta} \; .
\end{eqnarray}
The contribution from Vertex Operator \#9 involves the other contraction of
the propagator. The pole at $D\!=\!2$ cancels when the two are summed,
\begin{equation}
\Bigl({\rm Graviton\ Loop}\Bigr)^{\alpha\beta}_{3+9} = \frac{i \kappa H^D a^D
}{(4\pi)^{\frac{D}2}} \frac{\Gamma(D-1)}{\Gamma(\frac{D}2)} \left\{\Bigl(
\frac{D\!-\!1}{D\!-\!3}\Bigr) - \frac12 (D\!-\!1)^2\right\} \eta^{\alpha\beta}
\; .
\end{equation}

The entire graviton loop sums to,
\begin{eqnarray}
\lefteqn{\Bigl({\rm Graviton\ Loop}\Bigr)^{\alpha\beta} = \frac{i \kappa H^D 
a^D}{(4\pi)^{\frac{D}2}} \frac{\Gamma(D-1)}{\Gamma(\frac{D}2)} \left\{\frac12
D (D\!+\!1) \, \overline{\eta}^{\alpha\beta} - \frac12 (D\!-\!1) D \, 
\delta_0^{\alpha} \delta_0^{\beta} \right. } \nonumber \\
& & \hspace{-.3cm} \left. + \left[\Bigl(\frac{D\!-\!1}{D\!-\!3}\Bigr) - \frac12
(D\!-\!1)^2 (D\!+\!1) + \frac12 (D\!-\!1) + \frac18 (D\!-\!4) (D\!-\!1)^2
\right] \eta^{\alpha\beta} \right\} . \qquad \label{gravloop}
\end{eqnarray}
Note that all exotic denominators have canceled, save for a lone factor of
$1/(D-3)$ from Vertex Operator \#3. Because this multiplies $\eta^{\alpha
\beta}$ it can be absorbed into a harmless renormalization of the
cosmological constant.

The contributions from the ghost loop are comparatively simple to evaluate.
They are listed in Table 4. Their sum is,
\begin{equation}
\Bigl({\rm Ghost\ Loop}\Bigr)^{\alpha\beta} = \frac{i \kappa H^D a^D}{(4\pi)^{
\frac{D}2}} \frac{\Gamma(D-1)}{\Gamma(\frac{D}2)} \left\{\frac12 (D\!-\!1) \, 
\eta^{\alpha\beta} - D \, \overline{\eta}^{\alpha\beta} \right\} . 
\label{ghostloop}
\end{equation}
Adding (\ref{gravloop}) and (\ref{ghostloop}) gives the total for the two
primitive graphs of Fig.~1,
\begin{eqnarray}
\lefteqn{\Bigl({\rm Primitive}\Bigr)^{\alpha\beta} = \frac{i \kappa H^D a^D}{
(4\pi)^{\frac{D}2}} \frac{\Gamma(D-1)}{\Gamma(\frac{D}2)} \left\{\Bigl(
\frac{D\!-\!1}{D\!-\!3}\Bigr) \right. } \nonumber \\
& & \hspace{3cm} \left. - \frac12 (D \!-\! 2) (D\!-\!1) (D\!+\!1) + \frac18 
(D\!-\!4) (D\!-\!1)^2 \right\} \eta^{\alpha\beta} \; . \qquad \label{prim}
\end{eqnarray}
Because all noncovariant terms have canceled the entire one loop result can
be absorbed into a renormalization of the cosmological constant,
\begin{equation}
\delta\Lambda \!=\! \frac{\kappa^2 H^D}{(4\pi)^{\frac{D}2}} \frac{\Gamma(D-2)}{
\Gamma(\frac{D}2)} \!\left\{\!2\Bigl(\frac{D\!-\!1}{D\!-\!3}\Bigr) \!-\! (D 
\!-\! 2) (D\!-\!1) (D\!+\!1) \!+\! \frac14 (D\!-\!4) (D\!-\!1)^2 \!\right\}.
\end{equation}
In fact it {\it must} be so absorbed if our renormalization condition is that
the universe begins inflation with Hubble constant $H$. Hence our result for
the one loop 1PI 1-point function is zero!
 
\begin{table}

\vbox{\tabskip=0pt \offinterlineskip
\def\tablerule{\noalign{\hrule}}
\halign to450pt {\strut#& \vrule#\tabskip=1em plus2em& 
\hfil#& \vrule#& \hfil#\hfil& \vrule#& \hfil#& \vrule#& \hfil#\hfil& 
\vrule#\tabskip=0pt\cr
\tablerule
\omit&height4pt&\omit&&\omit&&\omit&&\omit&\cr
&&\omit\hidewidth \# 
&&\omit\hidewidth {\rm Vertex Contribution}\hidewidth&& 
\omit\hidewidth \#\hidewidth&& 
\omit\hidewidth {\rm Vertex Contribution}
\hidewidth&\cr
\omit&height4pt&\omit&&\omit&&\omit&&\omit&\cr
\tablerule
\omit&height2pt&\omit&&\omit&&\omit&&\omit&\cr
&& 1 && $-{\scriptstyle (D-1)} \overline{\eta}^{\alpha\beta} - 
\delta^{\alpha}_0 \delta^{\beta}_0$
&& 6 && $0$ & \cr
\omit&height2pt&\omit&&\omit&&\omit&&\omit&\cr
\tablerule
\omit&height2pt&\omit&&\omit&&\omit&&\omit&\cr
&& 2 && $-\frac{(D-1)}{D} \overline{\eta}^{\alpha\beta} - \frac1{D}
\delta^{\alpha}_0 \delta^{\beta}_0$
&& 7 && $0$ & \cr
\omit&height2pt&\omit&&\omit&&\omit&&\omit&\cr
\tablerule
\omit&height2pt&\omit&&\omit&&\omit&&\omit&\cr
&& 3 && $0$
&& 8 && $0$ & \cr
\omit&height2pt&\omit&&\omit&&\omit&&\omit&\cr
\tablerule
\omit&height2pt&\omit&&\omit&&\omit&&\omit&\cr
&& 4 && $0$
&& 9 && $\frac12 {\scriptstyle (D-1)} \eta^{\alpha\beta}$ & \cr
\omit&height2pt&\omit&&\omit&&\omit&&\omit&\cr
\tablerule
\omit&height2pt&\omit&&\omit&&\omit&&\omit&\cr
&& 5 && $\frac{(D-1)}{D} \overline{\eta}^{\alpha\beta} + \frac1{D}
\delta^{\alpha}_0 \delta^{\beta}_0$
&& 10 && $-\eta^{\alpha\beta}$ & \cr
\omit&height2pt&\omit&&\omit&&\omit&&\omit&\cr
\tablerule}}

\caption{Contributions from the ghost loop with an overall factor of 
$\frac{i \kappa H^D a^D}{(4\pi)^{\frac{D}2}} \frac{\Gamma(D-1)}{\Gamma(\frac{D
}2)}$ removed.}

\end{table}

\section{Discussion}

We have used dimensional regularization to compute the 1PI graviton 1-point
function at one loop order about a locally de Sitter background. Like other
computations, our result can be expressed as a finite shift in the 
background cosmological constant. We can write this as the negative of
the cosmological counterterm $\delta \Lambda$ that would be needed to cancel 
the effect and enforce the elementary consistency condition that the
universe begins inflation at the background Hubble constant. Recall that the
Newtonian expectation is \cite{TW0},
\begin{equation}
-\delta \Lambda_{{\rm Newt}} = +\frac1{16}  \frac{\kappa^2 H^4}{\pi^2}\; .
\end{equation}
Our result is a factor of 24 larger,
\begin{equation}
-\delta \Lambda_{{\rm TW} \atop {D=4}} = +\frac32 \frac{\kappa^2 H^4}{\pi^2} 
\; .
\end{equation}
By comparison, the recent result by Finelli, Marozzi, Venturi and Vacca is 
\cite{FMVV},
\begin{equation}
-\delta \Lambda_{{\rm FMVV}} = +\frac{361}{1920} \frac{\kappa^2 H^4}{\pi^2}\; .
\end{equation}
When one corrects for the normalization of the graviton, Ford's result is
\cite{LHF},
\begin{equation}
-\delta \Lambda_{{\rm Ford}} = - \frac{\kappa^2 H^4}{\pi^2}\; .
\end{equation}

The failure of any of these results to agree seems arise to from having 
sometimes computed different things and sometimes used different techniques. 
The Newtonian model derives from an estimate for just the infrared 
contributions under the assumption that each graviton polarization 
contributes to the vacuum energy like a massless, minimally coupled scalar. 
Ford's result is a direct computation of what gravitons contribute to the
vacuum, but only from infrared gravitons. By contrast, the result of Finelli, 
Marozzi, Venturi and Vacca includes ultraviolet gravitons, which can of course 
induce additional constant shifts in the vacuum energy. Our result also 
includes the full theory, but in a different gauge and with a different 
regularization. It has long been known that even the finite parts of 
counterterms can disagree in different gauges and with different 
regularization techniques. On the {\it physical} result everyone agrees: 
the one loop effect can be absorbed into a counterterm.

So there are good grounds for believing we have succeeded in dimensionally
regulating quantum gravity about de Sitter background. An obvious first 
application for this formalism is to re-compute the one loop graviton 
self-energy that was previously obtained using a cutoff on the co-moving 
3-momentum \cite{TW4}. One would then like to study the quantum-corrected,
linearized Einstein equations,
\begin{equation}
D_{\mu\nu}^{~~\rho\sigma} h_{\rho\sigma}(x) - \int d^4x' \Bigl[\mbox{}_{\mu\nu}
\Sigma^{\rho\sigma}\Bigr](x;x') h_{\rho\sigma}(x') = -\frac{\kappa}{2a^2}
T_{\mu\nu}(x) \; .
\end{equation}
The solution with zero stress tensor describes how one loop corrections
modify free gravitons. 

Solving the quantum-corrected, linearized Einstein equations with the stress 
tensor of a point mass would determine how one loop corrections affect the 
long range force law. In this regard it is interesting to note that there is 
no simple dimensional argument that one loop corrections must be negligible
at large distances the way they must be in flat space \cite{JFD1,JFD2}. In
de Sitter background the universal one loop factor of $\kappa^2$ can be 
balanced by a factor of $H^2$, rather than just the $1/r^2$ of flat space. 
Loop corrections can also acquire factors of the number of e-foldings since 
the onset of inflation. So it seems entirely possible for the long range 
force in de Sitter background to acquire a secular proportional correction 
of the form $\kappa^2 H^2 \ln(a)$, which could become nonperturbatively strong 
over a very long period of inflation.

Other obvious, and fairly simple, applications for the new formalism are
computing the quantum gravitational contributions to the one loop scalar 
self-mass-squared and to the one loop fermion self-energy. Both models
show enhanced quantum effects from scalar couplings \cite{BOW,PW3} so it 
is reasonable to expect enhanced effects from gravitons. These studies 
are on-going and results should be available soon.

A more complicated but timely application would be applying dimensional 
regularization to modified gravity models involving inverse powers
the Ricci scalar which have been invoked to explain the recent phase of 
cosmological acceleration \cite{CDTT}. A heroic computation of the one loop 
effective potential (as a function of constant curvature) has recently been 
carried out using generalized zeta function regularization \cite{CENOZ}. 
Because nonlinear functions of only the Ricci scalar would just change the 
scalar part of the graviton propagator it should not be prohibitively 
difficult to generalize our methods to these models.

\vspace{1cm}

\centerline{\bf Acknowledgements}

It is a pleasure to acknowledge conversations with A. O. Barvinsky,
F. Finelli and L. H. Ford. This work was partially supported by European 
Union grants FP-6-012679 and MRTN-CT-2004-512194, by NSF grant PHY-0244714, 
and by the Institute for Fundamental Theory at the University of Florida.


\begin{thebibliography}{99}

\bibitem{HV1} G.`t Hooft and M. J. G. Veltman, Nucl. Phys. {\bf B44} (1972) 
189.

\bibitem{BG} C. G. Bollini and J. J. Giambiagi, Nuovo Cim. {\bf B12} (1972) 20.

\bibitem{HV2} G. `t Hooft and M. J. G. Veltman, Ann. Poincare Phys. Theor.
{\bf A20} (1974) 69.

\bibitem{DN1} S. Deser and P. van Nieuwenhuizen, Phys. Rev. {\bf D10} (1974) 
411.

\bibitem{DN2} S. Deser and P. van Nieuwenhuizen, Phys. Rev. {\bf D10} (1974) 
401.

\bibitem{DTN} S. Deser, H. S. Tsao and P. van Nieuwenhuizen, Phys. Rev. {\bf 
D10} (1974) 3337.

\bibitem{GS} M. H. Goroff and A. Sagnotti, Nucl. Phys. {\bf B266} (1986) 709.

\bibitem{LHF} L. H. Ford, Phys. Rev. {\bf D31} (1985) 710.

\bibitem{FMVV} F. Finelli, G. Marozzi, G. P. Vacca and G. Venturi,
Phys. Rev. {\bf D71} (2005) 023522, gr-qc/0407101.

\bibitem{RPW1} R. P. Woodard, ``de Sitter Breaking in Field Theory,'' 
gr-qc/0408002.

\bibitem{TW0} N. C. Tsamis and R. P. Woodard, Ann. Phys. {\bf 267} (1998)
145, hep-ph/9712331.

\bibitem{TW1} N. C. Tsamis and R. P. Woodard, Ann. Phys. {\bf 253} (1997) 1,
hep-ph/9602317.

\bibitem{TW2} N. C. Tsamis and R. P. Woodard, Nucl. Phys. {\bf B474} (1996) 
235, hep-ph/9602315.

\bibitem{MC} V. F. Mukhanov and G. V. Chibisov, JETP Lett. {\bf 33} (1981) 532,
astro-ph/0303077.

\bibitem{AAS} A. A. Starobinski\u{\i}, JETP LEtt. {\bf 30} (1979) 682.

\bibitem{OW1} V. K. Onemli and R. P. Woodard, Class. Quant. Grav. {\bf 19} 
(2002) 4607, gr-qc/0204065.

\bibitem{OW2} V. K. Onemli and R. P. Woodard, Phys. Rev. {\bf D70} (2004) 
107301, gr-qc/0406098.

\bibitem{BOW} T. Brunier, V. K. Onemli and R. P. Woodard, Class. Quant. Grav.
{\bf 22} (2005) 59, gr-qc/0408080.

\bibitem{PTW1} T. Prokopec, O. Tornkvist and R. P. Woodard, Phys. Rev. Lett.
{\bf 89} (2002) 101301, astro-ph/0205331.

\bibitem{PTW2} T. Prokopec, O. Tornkvist and R. P. Woodard, Ann. Phys. 
{\bf 303} (2003) 251, gr-qc/0205130.

\bibitem{PW1} T. Prokopec and R. P. Woodard, Ann. Phys. {\bf 312} (2004) 1,
gr-qc/0310056.

\bibitem{DDPT} A. C. Davis, K. Dimopoulos, T. Prokopec and O. Tornkvist,
Phys. Lett. {\bf B501} (2001) 165, astro-ph/0007214.

\bibitem{DPTD} K. Dimopoulos, T. Prokopec, O. Tornkvist and A. C. Davis,
Phys. Rev. {\bf D65} (2002) 063505, astro-ph/0108093.

\bibitem{PW2} T. Prokopec and R. P. Woodard, Am. J. Phys. {\bf 72} (2004) 60,
astro-ph/0303358.

\bibitem{PW3} T. Prokopec and R. P. Woodard, JHEP {\bf 0310} (2003) 059,
astro-ph/0309593.

\bibitem{DW} L. D. Duffy and R. P. Woodard, Phys. Rev. {\bf D72} (2005)
024023, hep-ph/0505156.

\bibitem{TW3} N. C. Tsamis and R. P. Woodard, Commun. Math. Phys. {\bf 162} 
(1994) 217.

\bibitem{BD} N. D. Birrell and P. C. W. Davies, {\it Quantum Fields in Curved 
Space} (Cambridge University Press, Cambridge, 1982).

\bibitem{AF} B. Allen and A. Folacci, Phys.Rev. {\bf D35} (1987) 3771.

\bibitem{BA} B. Allen, Phys. Rev. {\bf D32} (1985) 3136.

\bibitem{TW4} N. C. Tsamis and R. P. Woodard, Phys. Rev. {\bf D54} 
(1996) 2621, hep-ph/9602317.

\bibitem{JFD1} J. F. Donoghue, Phys. Rev. Lett. {\bf 72} (1994) 2996,
gr-qc/9310024.

\bibitem{JFD2} J. F. Donoghue, Phys. Rev. {\bf D50} (1994) 3874, 
gr-qc/9405057.

\bibitem{CDTT} S. M. Carroll, V. Duvvuri, M. Trodden and M. S. Turner,
Phys. Rev. {\bf D70} (2004) 043528, astro-ph/0306438.

\bibitem{CENOZ} G. Cognola, E. Elizalde, S. Nojiri, S. D. Odintsov and
S. Zerbini, JCAP {\bf 0502} (2005) 010, hep-th/0501096.

\end{thebibliography}
\end{document}